\documentclass[aps,twocolumn,prd,nofootinbib,superscriptaddress]{revtex4-1}

\usepackage{geometry}
\usepackage{graphicx}
\usepackage[caption=false]{subfig}
\usepackage{mathrsfs}

\usepackage{amsmath, bm}
\usepackage{hhline}
\usepackage{xcolor}

\usepackage{verbatim}
\usepackage{dcolumn}
\usepackage{color}
\usepackage{xspace}
\usepackage{url}
\usepackage{float}
\usepackage{multirow}
\usepackage{hyperref}
\usepackage[dvipsnames]{xcolor}

\newcommand{\secref}[1]{Section~\ref{#1}}
\newcommand{\appendref}[1]{Appendix~\ref{#1}}
\newcommand{\figref}[1]{Figure~\ref{#1}}
\newcommand{\tabref}[1]{Tab.~\ref{#1}}
\newcommand{\eqnref}[1]{eq.~\eqref{#1}}                       

\newcommand{\Figref}[1]{Figure~\ref{#1}}
\newcommand{\Tabref}[1]{Table~\ref{#1}}
\newcommand{\Eqnref}[1]{Equation~\eqref{#1}}

\newcommand{\Lmarg}{\mathcal{L}_{\rm{marg}}(\pmb{\lambda})}
\newcommand{\lnLmarg}{\ln \mathcal{L}_{\rm{marg}}}
\newcommand{\lnL}{\ln \mathcal{L}}
\newcommand{\mc}{{\cal M}_{\rm{c}}}

\newcommand{\chieff}{\chi_{\rm eff}}
\newcommand{\frefnr}{f_{\rm ref}^{\rm NR}}
\newcommand{\fref}{f_{\rm ref}}
\newcommand{\flow}{f_{\rm low}}
\newcommand{\bayesecc}{\log_{10}\mathcal{B}_{\rm{ecc}}}
\newcommand{\bayesprec}{\log_{10}\mathcal{B}_{\rm{prec}}}

\def\RIT{Center for Computational Relativity and Gravitation, Rochester Institute of Technology, Rochester, New York 14623, USA}
\def\UNICAL{Dipartimento di Fisica, Universit\`a della Calabria, Arcavacata di Rende (CS), 87036, Italy}

\begin{document}
\title{Disentangling Spin-induced Precession vs. Eccentricity in GW190521 and GW231123}
\author{Patricia McMillin}
\affiliation{\RIT}
\author{Giuseppe Ficarra}
\affiliation{\UNICAL}
\author{Carlos O. Lousto}
\affiliation{\RIT}
\author{Richard O'Shaughnessy}
\affiliation{\RIT}
\begin{abstract}
We use the more than 2000 full numerical-relativity simulations 
included in the fifth release of the RIT binary-black-hole waveform 
catalog to perform Bayesian parameter estimation, via the rapid iterative 
fitting algorithm (RIFT), for the gravitational-wave 
events GW190521 and GW231123. We revisit GW190521, which has previously been 
identified as potentially possessing large eccentricity upon entering the 
detector band at $20~\mathrm{Hz}$, and confirm this finding: the highest 
likelihood is obtained from waveforms that are simultaneously 
eccentric and spin-precessing, yielding $e_{20} =0.207_{-0.147}^{+0.192}$ 
and $\chi_{1,\perp}=0.717_{-0.406}^{+0.194}$. 
Applying the same simulation pool to GW231123, we likewise find a preference 
for spin precession where at 20 Hz $\chi_{1,\perp}=0.678_{-0.462}^{+0.218}$, but in this case with negligible eccentricity at both 
$10~\mathrm{Hz}$ and $20~\mathrm{Hz}$, with 
$e_{10}=0.097_{-0.073}^{+0.124}$ and $e_{20}=0.056_{-0.045}^{+0.059}$.
For comparison, both events are also analyzed using exclusively 
non-precessing (eccentric and quasi-circular) simulations, which yield lower 
likelihoods in each case. These results indicate that our approach 
is capable of discriminating between the effects of eccentricity and 
spin precession in current gravitational-wave signals.
The singled out top peak likelihood NR simulations give estimates of the 
recoil velocities of the merged final holes of 3965km/s for GW190521, 
and of 2059km/s for GW231123.
\end{abstract}
\maketitle

\section{Introduction}
\label{s:intro}
The International Gravitational Wave Observatory Network (IGWN) of 
ground-based gravitational-wave (GW) detectors---comprised of Advanced LIGO 
\cite{2015CQGra..32g4001L}, Virgo \cite{virgo}, and, more recently, KAGRA 
\cite{kagra}---continues to detect coalescing compact binaries 
\cite{DiscoveryPaper150914,LIGO-O1-BBH,LIGO-GW170817-bns,LIGO-O3-NSBH,LIGO-O3-O3a_final-catalog,LIGO-O3-O3b-catalog,LIGO-O4-O4a_catalog,LIGO-O4-O4b_catalog}.
The astrophysical properties of these sources are inferred by comparing the 
observed GW signals with theoretical predictions of the radiation emitted by 
merging binary black holes, obtained either from detailed numerical-relativity 
(NR) calculations or from phenomenological waveform models.

Most compact binaries entering the LIGO-Virgo-KAGRA (LVK) sensitivity band 
at $10~\mathrm{Hz}$ are expected to be in quasi-circular orbits. Systems formed 
through isolated binary evolution typically originate at separations large 
enough that their merger timescales allow the gravitational radiation reaction to 
circularize the orbit well before coalescence 
\cite{Bethe_1998, Belczynski_2002, Stevenson_2017}. 
However, recent studies have suggested that a subset of compact binaries may 
retain non-negligible orbital eccentricity at merger. Such systems may signal 
an alternative formation pathway known as dynamical assembly. Binaries formed 
through this channel arise from frequent chance encounters in densely 
populated stellar environments \cite{Mandel_2010}, including globular 
clusters \cite{Sadowski_2008,Rodriguez_2018a,Zevin_2019}, nuclear star 
clusters \cite{Fragione_2020}, and galactic centers \cite{Samsing_2014}, 
where multi-black-hole interactions can be enhanced 
\cite{Campanelli:2007ea,Lousto:2007rj,Bai:2011za,Imbrogno:2021xrh,Ficarra:2023zjc,Ficarra:2024jen,Bamber:2025gxj,Heinze:2025usf,Wu:2026hth,Cardoso:2026ugm}.
In these settings, previously evolved compact objects may undergo one or 
several gravitational interactions that lead to the formation and rapid 
merger of bound pairs. Although such binaries are typically short-lived, 
they can retain a non-negligible eccentricity at merger that is potentially 
measurable with current and future GW detectors \cite{Rodriguez_2018b}.

Dynamical formation channels can also influence the spins of a binary black hole
(BBH), as significant spin-induced precession is more likely to be produced 
\cite{Rodriguez_2016_BBHspin} in dynamical channels, whereas isolated channels 
struggle to produce highly precessing systems 
\cite{Gerosa_2018,Steinle_2021,Kalogera_2000}. In particular, high mass BBHs which 
have component masses that lie in the pair-instability mass gap 
($M^{\rm source}\approx 60-130 M_\odot$) require formation channels such as prior 
stellar mergers \cite{Di_Carlo_PI_BBH_2020,Kremer_2020_PIBBH,Renzo_2020_PIBBH}, 
hierarchical black hole (BH) mergers \cite{Gerosa_2021_hierarchBBH}, or accretion 
driven mergers \cite{McKernan_2012_agnBBH}. These channels tend to form highly 
spinning black holes with a dimensionless spin magnitude of $\chi \sim 0.7$ 
\cite{Gerosa_2021_hierarchBBH} where the 
inferred remnant spin may be even higher. Investigating eccentricity and 
spin-precession in parameter estimation simultaneously presents a few difficulties, 
due to the degeneracy between the two effects 
\cite{tibrewal2026_ecc_prec,Romero_Shaw_2023,Divyajyoti_2024,Divyajyoti_2026}, as well 
as the exploration of a larger parameter space. Waveform models 
incorporating both eccentricity and spin precession are relatively recent developments 
and the available models are generally slower at generating waveforms than their 
eccentric or precessing only counterparts. However, numerical relativity simulations 
with both effects are numerous, allowing for the simultaneous study of both effects 
across the entire parameter space.

A growing body of work 
\cite{Romero_Shaw_2019,RomeroShaw_2020,RomeroShaw_2021,
Romero_Shaw_2022,romero-shaw_2025,gupte_2025,Iglesias_2024,Gayathri:2020coq,Gamba_2022,
malagon_2026_RIFTecc,xu2025_o4a_ecc,Planas_2025_eccvsQC,Planas_2025_gw200105,
jan2026gw200105,Kacanja_2025_gw200105,Morras_2026_gw200105} 
has identified additional GW events potentially produced by binaries entering 
the LVK sensitivity band with non-negligible orbital eccentricity. 
Candidates warranting further investigation---noting that this is not a complete 
list---include GW190521, GW190620, GW191109, GW200208\_22, GW190701, GW200129, 
GW200105, and GW231123. This work focuses on GW190521 and GW231123 for 
their potential to be both eccentric and precessing.

Gayathri {\it et al.} \cite{Gayathri:2020coq} demonstrated that GW190521 
is most consistent with a highly eccentric black-hole merger. They generated 
611 targeted numerical-relativity simulations of eccentric binaries, 
yielding an effective set of $\sim 6\times 10^4$ gravitational waveforms 
spanning a broad range of total masses---substantially exceeding the 
high-eccentricity coverage previously available. By comparing these NR 
simulations to the observed data using the rapid iterative fitting algorithm 
RIFT \cite{rift}, they concluded that GW190521 is best described by a highly 
eccentric, precessing model with $e \sim 0.7$ at the beginning of the simulations. 
All inferred properties of GW190521 point to an origin involving repeated 
gravitational capture of black holes, making it one of the first LIGO-Virgo 
discoveries whose formation channel has been tentatively identified.

GW190521 was also analyzed by Gamba {\it et al.} \cite{Gamba_2022} under the 
hypothesis that the binary components are on hyperbolic orbits due to 
dynamical capture. They use the hyperbolic, and spin precessing versions of
the TEOBResumS waveform to perform Bayesian parameter estimation and found that 
GW190521 favored a dynamical capture scenario with hyperbolic orbits 
over a quasi-circular merger. Studies on GW190521 
\cite{Gamba_2022,Gayathri:2020coq,siegel_2023} 
agree that this event is the result of a dynamical formation scenario, even if they 
propose differing exact scenarios.

Malagon \& O'Shaughnessy \cite{malagon_2026_RIFTecc} re-analyzed 162 GW events 
identified in the LVK O3 \cite{LIGO-O3-O3b-catalog} and O4a \cite{LIGO-O4-O4a_catalog}
catalogs with two independent eccentric spin-aligned waveform models (SEOBNRv5EHM and 
TEOBResumS-Dalí) using RIFT \cite{rift,Wysocki_2019, Wofford_2022,Wagner_2025} for 
Bayesian parameter estimation. They find evidence for eccentricity, quantified by a 
$\log_{10}$ Bayes factor above 0.1, in three events: GW200129\_065458, GW231001\_140220, and 
GW231123\_135430 (referred to as GW231123 henceforth). In the analysis using 
TEOBResumS-Dalí, GW231123 has an eccentricity of 0.45 at 13.33 Hz which agrees 
with the spin-aligned results of Jan {\it et al.} \cite{jan2026_gw231123}. 
However, the waveform models used in \cite{malagon_2026_RIFTecc} do not 
simultaneously include eccentricity and spin precession thus lack the ability to 
disentangle the two effects in their analysis. This limitation is evident in the expanded analysis in 
Jan {\it et al.} \cite{jan2026_gw231123}, which uses both nonprecessing and precessing 
versions of TEOBResumS-Dalí, and finds that spin precession is strongly favored over eccentricity.

Abac {\it et al.} \cite{GW231123_special} analyzed GW231123 with 5 inspiral-merger-ringdown 
models which incorporate precessing spins, where the model NRSur7dq4
\cite{Varma:2019csw} fully interpolates two-spin precessing systems in the precessing 
sector. They find strong support for component spins above $>0.8$ which lies in a 
region where no waveform model is reliably calibrated to NR waveforms from precessing 
binaries with spins above this threshold. Due to this fact they report their results 
as a combined posterior of all 5 models in an effort to reduce the waveform
uncertainties. The effective spin precession is found to be $\chi_{\rm p}\sim0.77$ 
with consistently large Bayes factors ($10^3:1$ to $10^8:1$) across waveforms in 
favor of the spin precessing hypothesis over the spin-aligned hypothesis. However,
the precession SNR $\rho_{\rm p}$, which should have $\rho_{\rm p}<2.1$ in the absence 
of precession for 90\% of cases, is found to be $\rho_{\rm p}=2.0_{-1.2}^{+5.2}$ thus 
precession can not be confidently claimed for the analysis done in 
\cite{GW231123_special}. While Abac {\it et al.} does not conduct a full analysis 
with eccentricity included as a parameter, they discuss the possibility citing that 
the likely formation scenario of at least one BH lying in the pair-instability mass 
gap $\approx 60-130 M_\odot$ suggests that eccentricity in this event is 
astrophysically unlikely. 

Xu {\it et al.} \cite{xu2025_o4a_ecc} reanalyze 84 GW events from GWTC-4.0 
\cite{LIGO-O4-O4a_catalog} using 3 waveforms from the IMRPhenom family that cover the 
quasicircular precessing case in both the frequency- and time-domain, and the 
nonprecessing eccentric case in the time-domain. Their study focuses on waveform 
systematic effects on the inferred parameters of a given event due to the choice of 
waveform. They find in the quasicircular analysis of GW231123 that the event
displays significant systematic differences between waveforms due to the event having 
very high detector frame total mass ($M>200 M_\odot$), large positive $\chieff$, and 
$\chi_{\rm p}$ distinct from the prior towards large values---which in this parameter 
space subdominant modes, ringdown modeling and spin-precessing modeling play key roles 
in parameter inference leading to larger systematic effects. This finding agrees with 
the work done in references \cite{siegel_2026,Bini_2026} as both studies find that 
GW231123 may have significant waveform systematic effects.
The eccentric analysis of GW231123 in \cite{xu2025_o4a_ecc} found that the eccentric 
spin-aligned hypothesis is preferred over the quasicircular aligned spin hypothesis, 
where the eccentricity at 10 Hz was found to be $e_{10\rm Hz}=0.57_{-0.05}^{+0.03}$, 
noting that the waveform model used is only accurate up to initial eccentricity of 
$e=0.4$. However when they compare the eccentric aligned spin case to the 
quasicircular precessing case the evidence vanishes as the Bayes factor is large and 
negative. Xu {\it et al.} conclude for GW231123 the inclusion of spin precession 
alongside eccentricity is vitally important for disentangling the two effects for 
this event.

In this work, we extend and refine previous results \cite{McMillin:2025hof} 
by exploiting numerical-relativity simulations from the RIT catalog 
\cite{RIT_NR1,RIT_NR2,RIT_NR3,RIT_NR4,Ficarra:2026cej} to probe high-likelihood 
regions of parameter space. We revisit GW190521 and perform new parameter 
estimation with RIFT \cite{rift}, using the expanded set of eccentric 
simulations included in the fifth RIT waveform catalog \cite{Ficarra:2026cej}, 
in order to better constrain the potentially non-vanishing eccentricity 
$e_{20}$ at $20~\mathrm{Hz}$. In addition, we apply the fifth RIT catalog to 
GW231123 to demonstrate that our parameter-space coverage enables robust 
parameter estimates of an event that lies in a region where large waveform model 
systematics are present due to the very high mass, high spins, and possible 
presence of eccentricity and spin-induced precession.
For this event, we find that the highest-likelihood simulations correspond to 
strongly precessing, quasi-circular configurations.

This paper is organized as follows. In \secref{s:methods}, we provide an 
overview of our methodology: \secref{sec:FN} describes the numerical 
techniques used to produce our simulations, \secref{ss:rift} reviews the 
RIFT framework as applied in this study, \secref{ss:NRsims} introduces 
the bank of NR simulations employed, and \secref{ss:analysis} details the 
analysis settings. In \secref{s:events}, we present the results of 
NR-based parameter inference for GW190521 and GW231123. Finally, in 
\secref{s:conclusion}, we summarize our findings and discuss directions 
for future development and application.

\section{Methods}
\label{s:methods}
\subsection{Full Numerical Techniques}
\label{sec:FN}
To carry out the full numerical-relativity simulations of binary black holes, 
we employ the {\scriptsize LazEv} code \cite{Zlochower:2005bj}, which uses 8th-order spatial 
finite differences \cite{Lousto:2007rj}, 4th-order Runge-Kutta time 
integration, and a reduced \cite{Zlochower:2012fk} Courant factor 
$c = dt/dx = 1/4$.

Initial data for the binary black holes are constructed using the puncture 
approach \cite{Brandt97b} in conjunction with the {\scriptsize TwoPunctures} code 
\cite{Ansorg:2004ds}. Apparent horizons are located during the evolution 
with {\scriptsize AHFinderDirect} \cite{Thornburg2003:AH-finding}, and horizon masses 
are computed from the horizon area $A_H$. The magnitudes of the horizon 
spins $S_H$ are evaluated using the ``isolated horizon'' algorithm 
\cite{Ashtekar:2004cn}, as implemented in Ref.~\cite{Campanelli:2006fy}.

Mesh refinement is handled by the {\sc Carpet} driver 
\cite{Schnetter-etal-03b}, which tracks the motion of the black holes 
across the numerical domain. {\sc Carpet} implements a ``moving boxes'' 
refinement scheme, in which refined grids of fixed size are centered on 
the coordinate locations of the black holes and follow their trajectories 
throughout the simulation.

For typical simulations, the outermost refinement box extends to $\pm 400M$. 
The coarsest grid contains XXX points between $0$ and $400M$ (e.g., n100 
has $100$ points), corresponding to a grid spacing of $400/\mathrm{XXX}$ 
on this level. The resolution in the wave zone is $100M/\mathrm{XXX}$ 
(e.g., n100 corresponds to $M/1.00$, n120 to $M/1.2$, and n144 to $M/1.44$), 
with the intermediate levels adjusted accordingly. For comparable-mass, 
non-spinning binaries, the finest grid surrounding one of the black holes 
($m_1$) is fixed at $\pm 0.65M$ in size and corresponds to the 9th 
refinement level, so the grid spacing at the highest refinement level is 
$400/\mathrm{XXX}/2^8$. For small mass-ratio binaries, additional internal 
refinement levels are introduced progressively \cite{Lousto:2020tnb}. 
We adopt units in which $M = m_1^H + m_2^H$, the sum of the horizon masses.

Gravitational radiation is extracted from the simulations using Eqs.~(22) 
and (23) of Ref.~\cite{Campanelli:1998jv} for the radiated energy and linear 
momentum, respectively, together with the expressions in 
Ref.~\cite{Lousto:2007mh} for the radiated angular momentum. All quantities 
are computed from the extracted Weyl scalar $\Psi_4$ at the observer 
location $R_{\rm obs} = 113M$. Extrapolation to future null infinity is 
performed using the perturbative formulas of Ref.~\cite{Nakano:2015pta}. 
This extraction radius provides sufficient accuracy for the low- to 
moderate-eccentricity regime considered in this paper; a detailed 
investigation is presented in Ref.~\cite{Ficarra:2024nro}.

Eccentric binaries are evolved starting from an initial separation 
corresponding to the apoapsis, $r_a = a_r (1 + e_r)$, where the initial 
tangential linear momentum $P_t$ is reduced by a factor $1 - f$ relative 
to the quasi-circular value $P_c$ at the same apoapsis, namely 
$P_t = P_c (1 - f)$. The relation $e_r(f)$ has been computed to 3.5 
post-Newtonian (PN) order in Ref.~\cite{Ciarfella:2022hfy}, and this 
prescription has been adopted since the fifth release of the RIT BBH 
catalog \cite{Ficarra:2026cej}.

Because NR simulations are invariant under rescaling of the total mass 
$M$, the physical scale is fixed only after matching to a specific signal. 
The eccentricity at a chosen reference frequency, such as $\fref=20~\mathrm{Hz}$ 
at apoapsis, can therefore be evaluated {\it a posteriori} from its 
reference NR value $\frefnr$; see Fig. 5 of \cite{McMillin:2025hof} for 
details on the NR reference frequency. We note that the apoapsis frequency $f_a$ and the 
periapsis frequency $f_p$ define the mean orbital frequency 
$f_{\rm avg} = (f_a + f_p)/2$. To this end, we use the lowest-order 
post-Newtonian formula for the eccentricity evolution given in 
Eq.~(5.11) of Ref.~\cite{Peters:1964zz}, yielding
\begin{equation}\label{eq:e20}
\begin{split}
    e_{\fref} = \; & e_{\rm NR} \left(\frac{\frefnr}{\fref}\right)^{19/18} \\
    \times \Biggl[ 1 + & \frac{19}{9}\, e_{\rm NR}\left( 1 - \left(\frac{\frefnr}{\fref}\right)^{19/18} \right)
    \Biggr],
\end{split}
\end{equation}
which is valid for small eccentricities in the post-Newtonian inspiral 
regime, where $\fref$ is the reference frequency that eccentricity is evaluated at. 
We note that this conversion to $e_{20}$ or $e_{10}$ is carried out after 
both stages of RIFT are complete but before the final posteriors are generated, since it requires the rescaled mass 
of each simulation.

\subsection{RIFT}
\label{ss:rift}
A merging compact binary is fully characterized by its intrinsic and 
extrinsic parameters. The intrinsic parameters $\pmb{\lambda}$ comprise the 
component masses, component spins, eccentricity, and any matter degrees of 
freedom. The seven extrinsic parameters $\pmb{\theta}$ specify the spacetime 
location and orientation of the system: right ascension, declination, 
luminosity distance, coalescence time, inclination, orbital phase, and 
polarization. Throughout this work, masses are expressed in solar units in 
the detector frame, and dimensionless aligned spins are reported through their
Cartesian components along the orbital angular momentum, $\chi_{i,z}$.

RIFT \cite{rift, Wysocki_2019, Wofford_2022,Wagner_2025} is an iterative algorithm, 
consisting of two stages, that estimates the intrinsic and extrinsic 
parameters of a binary source by comparing gravitational-wave data $d$ 
with predicted signals $h(\pmb{\lambda}, \pmb{\theta})$. 

In the first stage, for each parameter point $\pmb{\lambda}_\alpha$ drawn 
from a proposed ``grid'' $\alpha = 1, 2, \ldots, N$ of candidate intrinsic 
parameters, RIFT evaluates---in parallel---the marginal likelihood
\begin{equation}
    \Lmarg = \int \mathcal{L}(\pmb{\lambda}, \pmb{\theta})\, p(\pmb{\theta})\, d\pmb{\theta},
\end{equation}
where $\mathcal{L}(\pmb{\lambda}, \pmb{\theta})$ is the likelihood of the 
gravitational-wave signal in a multi-detector network, including the 
detector response. This stage, referred to as Integrate Likelihood 
Extrinsic (ILE), produces point estimates of $\ln \Lmarg$, which are then 
interpolated over the intrinsic parameter space using either Gaussian 
process regression or random forests. We refer the reader to 
Refs.~\cite{rift, Wysocki_2019, Wofford_2022,Wagner_2025} for further details.

\subsubsection{NR-Based RIFT}
\label{sss:NRRIFT}
In the second stage---Construct Intrinsic Posterior (CIP)---RIFT 
first constructs an approximation of $\mathcal{L}(\pmb{\lambda})$ 
from the set of evaluations $\{(\pmb{\lambda}_\alpha, \mathcal{L}_\alpha)\}$. 
From this approximation, 
it generates a full posterior distribution over the intrinsic parameters,
\begin{equation}
    p_{\rm post}(\pmb{\lambda}) = \frac{\Lmarg\, p(\pmb{\lambda})}{\int d\pmb{\lambda}\, \Lmarg\, p(\pmb{\lambda})}\,,
\end{equation}
where $p(\pmb{\lambda})$ is the prior on the intrinsic parameters.
The posterior is then sampled with adaptive Monte Carlo techniques to produce 
a new grid for the next iteration. When numerical-relativity (NR) 
simulations are used as grid points we must modify the MC fit to accommodate 
the fixed intrinsic parameters of NR simulations. Each NR simulation corresponds 
to a fixed combination of the intrinsic 
parameters $\pmb{\lambda}$---mass ratio, spin components, and 
eccentricity---but can be rescaled to any value of the total mass. 
Consequently, the second stage of RIFT employs a one-dimensional fit to 
construct a refinement grid in the vicinity of the peak $\ln \Lmarg$, 
adaptively exploring only the total-mass parameter space for each combination of the 
remaining intrinsic parameters available in the NR catalog. Exploration 
in the other intrinsic parameters is therefore restricted to the discrete 
set of configurations present in the catalog. The evaluation points and 
associated weights in this integral correspond to weighted posterior samples, 
which are subsequently resampled into conventional, independent and identically
distributed posterior samples. For additional details on the 
technical foundations and performance of RIFT, see 
Refs.~\cite{rift, Wysocki_2019, Wofford_2022, Wagner_2025}.

\subsection{Simulations}
\label{ss:NRsims}
The fifth release of the RIT catalog of BBH simulations \cite{Ficarra:2026cej} 
consists of 2129 accurate simulations, 1021 in eccentric orbits with 
$0 < e \leq 1$, including 906 nonprecessing and 115 precessing; and 625 
nonprecessing and 483 precessing quasicircular/inspiraling binary systems with 
mass ratios $q = m_2/m_1$ in the range $1/128 \leq q \leq 1$ 
and individual spins up to $\chi_i = S_i/m_i^2 = 0.97$. The catalog also 
provides initial parameters of the binary, trajectory information, peak 
radiation, and final remnant black hole properties. The catalog includes all 
waveform modes $\ell \leq 4$ of the Weyl scalar $\psi_4$ and the strain $h$ up 
to $\ell=5$ for eccentric BBH (both 
extrapolated to null-infinity) and both are corrected for the center of mass 
displacement during inspiral and after merger \cite{RIT_NR4}. 
We use waveform modes $\ell \leq 4$ in all RIFT analyses.
We note here that even our extended bank of 
simulations does not have an even coverage of 
parameter space and contains sub-regions with few simulations which may affect the posteriors we generate.

\subsection{Analysis settings} 
\label{ss:analysis}
We analyze publicly available data from GWTC-2.1, and GWTC-4.0 
\cite{LIGO-O3-O3a_final-catalog,LIGO-O4-O4a_catalog} and use the noise power 
spectral density curves (PSDs) associated with each specific event as part of 
the public data release. We use public data from LIGO Hanford (H1), 
LIGO Livingston (L1), and Virgo (V1) for GW190521, and only H1 and L1 data 
for GW231123. For GW190521 we used a low-frequency cutoff of 
$\flow$ = 20 Hz, and for GW231123 a 20 Hz and a 10 Hz cutoff, where the 
latter is presented in \appendref{s:231123_e10}.  
The high frequency cutoff was 224 Hz for GW190521 and 448 Hz for GW231123. 
For details on the choice of high-frequency cutoff see Section V of 
\cite{LIGO-O3-O3a_final-catalog}, and Section 4.1 of \cite{GW231123_special} 
respectively for each event. In addition, we adopt a reference 
frequency $\fref$ used in the evaluation of eccentricity using \Eqnref{eq:e20} 
equal to the chosen $\flow$ for each event analysis.

We adopt a uniform prior for eccentricity over the range $e_{\fref} \in [0.0,0.9]$ 
for all events. We assume both spins are aligned with the orbital 
angular momentum and uniformly distributed in our prior range, where 
we construct our posteriors using each of these three priors 
$|\chi_{i,z}|<0.5$, $<0.9$, or $<0.99$. 
The more restrictive spin priors are used to identify 
the effects of excluding high spin simulations on our parameter estimates.
For analyses involving precessing simulations we have additional uniform priors 
on the remaining spin components $|\chi_{i,x}|<\chi_{\rm max}$ and 
$|\chi_{i,y}|< \chi_{\rm max}$ where $\chi_{\rm max}$ is the value chosen for the z 
component prior either $0.5, 0.9$ or $0.99$, the prior for the in-plane spin
components, $\chi_{i, \perp} \in [0,1]$ is constructed from the component spin prior. 
Where the in-plane spin components are defined as
\begin{equation}
    \chi_{i, \perp} = \left|\frac{\mathbf{S_i}}{m_i^2} \times \mathbf{\hat{L}}\right|.
\end{equation} 
All other extrinsic and intrinsic priors are conventional: for example, we use a 
uniform prior in the detector frame masses $m_{i}$, modulo upper and lower limits 
in $\mc$ and $q$; a Euclidean distance prior on $d_L$; and conventional uninformed 
priors on all extrinsic angles and event time. In RIFT's CIP stage the posteriors are constructed by sampling the cartesian spins $\chi_{i,j}$ where $i\in[1,2]$ and $j\in[x,y,z]$.

We utilize the nonprecessing and precessing simulations, both eccentric and non-eccentric, 
with symmetric mass ratio $\eta \in [0.02938,0.25]$ where $\eta=q/(1+q)^2$ (which 
corresponds to $q\in[0.03125,1.0]$). We use the RIT catalog \cite{Ficarra:2026cej} 
as an initial grid for NR-based PE where we rescale over the detector frame chirp 
mass range for each event, GW190521 has a chirp mass range of 
$\mc \in [20,178.7]\ M_\odot$, and GW231123 has $\mc \in [68.2,194.6]\ M_\odot$.

\section{Analysis of real events}
\label{s:events}
In this section we present the results of the eccentric reanalysis, 
including spin precession, of GW190521 and GW231123 using direct comparison 
to numerical relativity. These events were analyzed based on prior evidence or 
indications of eccentricity and/or spin-induced precession 
\cite{RomeroShaw_2020,Romero_Shaw_2022,RomeroShaw_2021,Gayathri:2020coq,
GW231123_special,malagon_2026_RIFTecc,jan2026_gw231123,Iglesias_2024,xu2025_o4a_ecc}.
In the discussion below we only employ comparisons to numerical 
relativity simulations.

When reporting the peak marginal likelihood points in Tables 
\ref{tab:190521_aligned_peakLnL},\ref{tab:190521_prec_peakLnL}, 
\ref{tab:231123_e20_aligned_peakLnL},\ref{tab:231123_e20_prec_peakLnL}, 
\ref{tab:231123_e10_aligned_peakLnL}, and \ref{tab:231123_e10_prec_peakLnL}, we 
perform a Gaussian process fit on each of the top 10 simulations 
presented. We rerun the ILE stage of RIFT for each simulation 
with the same settings as described in \secref{ss:analysis} but with the use 
of a denser initial grid in chirp mass with a spacing of 0.004 between points, where 
we perform the fit on the chirp mass region which returned a maximal 
likelihood. We take the chirp mass limits to be $\pm 5 \ M_\odot$ of the chirp 
mass of each maximal likelihood simulation, for example RIT:BBH:1794 in 
\Tabref{tab:190521_aligned_peakLnL} has $\mc=140.96\ M_\odot$ thus the new grid is 
over the region $\mc =140.96\ M_\odot\pm 5\ M_\odot$. The Gaussian process fit 
is performed on the output from the first stage of RIFT (ILE), we first perform a 
cut in marginal log-likelihood, by only considering the data points that fall 
within 3 of the maximum $\lnL$, then we fit the marginal likelihoods as a 
function of chirp mass. We employ an additive kernel \eqnref{eq:gp_kernel} composed 
of a white noise kernel with variance $\nu$ plus the product 
of a constant kernel with value $C=0.5$ and the radial basis function 
kernel with length scale, $\ell$, set at 2 times the standard deviation 
of chirp mass
\begin{equation}
    k(x_i,x_j) = \nu \delta_{ij} + C \exp\left( - \frac{|x_i-x_j|^2}{2\ell^2} \right),
    \label{eq:gp_kernel}
\end{equation}
where $x_i$ and $x_j$ are chirp-mass grid points from the ILE stage, the variance is 
$\nu = 6\times10^{-4}$ for GW190521 and $\nu = 1\times10^{-3}$ for GW231123.
We then compute the mean prediction of the GP fit and its 90\% confidence interval, 
then report the maximum $\lnL$ of the fit.

We report our parameter estimates by generating a kernel density estimate (KDE) on 
each of the one-dimensional posteriors where the maximum of each KDE is the reported 
median parameter along with the 90\% confidence interval on the 1D posterior. For 
additional details on this process see Appendix A of \cite{McMillin:2025hof}.

To quantify the support for eccentricity we adopt the Savage-Dickey (SD) density ratio 
used in \cite{malagon_2026_RIFTecc} to compute a Bayes factor between the quasi-
circular and eccentric hypotheses. Our analyses assume a uniform prior on 
eccentricity, $e_{\fref} \in [0,0.9]$, where the Bayes factor is calculated as
\begin{eqnarray}
    \label{eq:bayes}
    \bayesecc&&= -\log_{10}\left( \frac{p(e_{\fref}=0|d)}{p(e_{\fref}=0)}\right)\nonumber\\
    &&= - \log_{10}(0.9 \ p(e_{\fref}=0|d)),
\end{eqnarray}
where $p(e_{\fref}=0|d)$ is the posterior density at zero eccentricity for the 
quasicircular hypothesis given the data $d$. We use a boundary corrected KDE 
constructed by reflection at $e=0$ to estimate the posterior density. Evidence for 
the eccentric hypothesis is reflected by a positive $\log_{10}$ Bayes factor.

Constructing a Bayes factor for spin precession is not quite as simple. Since we 
sample the cartesian spins the analytic prior for $\chi_{1,\perp}$ goes to zero at 
$\chi_{1,\perp}=0$ and the SD density ratio approach breaks down. Instead 
we utilize RIFT's ability to directly integrate marginal likelihoods to quantify 
support for precession. We quantify support for spin precession with a Bayes factor 
constructed as the ratio of the evidences for the precessing hypothesis 
($\mathcal{H}_2$) and the nonprecessing hypothesis ($\mathcal{H}_1$)
\begin{equation}
    \label{eq:bayesprec}
    \bayesprec = \log_{10} \left( \frac{Z(d|\mathcal{H}_2)}{Z(d|\mathcal{H}_1)}\right)
\end{equation}
where the evidence is 
$Z(d|\mathcal{H})=\int d\pmb{\lambda}\, \mathcal{L}(\pmb{\lambda})\, p(\pmb{\lambda}|\mathcal{H})$. 
Evidence for the precessing hypothesis is reflected by a positive value of 
$\bayesprec$.

\subsection{GW190521}
\label{ss:gw190521}
The LVK collaboration identified in GWTC-2.1 \cite{LIGO-O3-O3a_final-catalog} 
GW190521 with SNR $14.3_{-0.4}^{+0.5}$ as an event with high-mass 
sources, where the component masses, mass ratio and chirp mass in the detector 
frame as reported in \cite{O3a_PE_data_release} are
$m_1=152.4_{-17.6}^{+31.7}\ M_\odot$, $m_2=89.7_{-52.2}^{+49.0}\ M_\odot$, 
$q=0.59_{-0.38}^{+0.33}$, and $\mc=101.0_{-33.7}^{+29.0}\ M_\odot$. This event also 
has support for negative effective spin where the reported effective spin from 
Table VIII of \cite{LIGO-O3-O3a_final-catalog} is $\chieff=-0.14_{-0.45}^{+0.50}$. 
GW190521 was further analyzed in \cite{LIGO-O3-GW190521-implications} under the 
assumption of quasicircular BBH coalescence, where the preferred results are from 
an analysis with the NRSur7dq4 \cite{Varma:2019csw} waveform. They find detector 
frame component masses, mass ratio, and chirp mass of $m_1=152_{-19}^{+32}\ M_\odot$, 
$m_2=120_{-32}^{+21}\ M_\odot$, $q=0.79_{-0.29}^{+0.19}$, and 
$\mc=117_{-14}^{+12}\ M_\odot$. In addition to the effective spin parameter 
$\chieff$, an effective spin-induced precession parameter $\chi_{\rm p}$ is 
included in their reported parameters as spins are not required to be aligned to 
the orbital angular momentum, they find $\chieff=0.008_{-0.36}^{+0.27}$ and 
$\chi_{\rm p}=0.68_{-0.37}^{+0.25}$. This expanded analysis 
\cite{LIGO-O3-GW190521-implications} concludes that this event has weak evidence for 
spin-induced orbital precession, and moderate inconclusive evidence for high spin components in the 
orbital plane.

\begin{figure*}[h!tbp]
    \includegraphics[width=1.45\columnwidth]{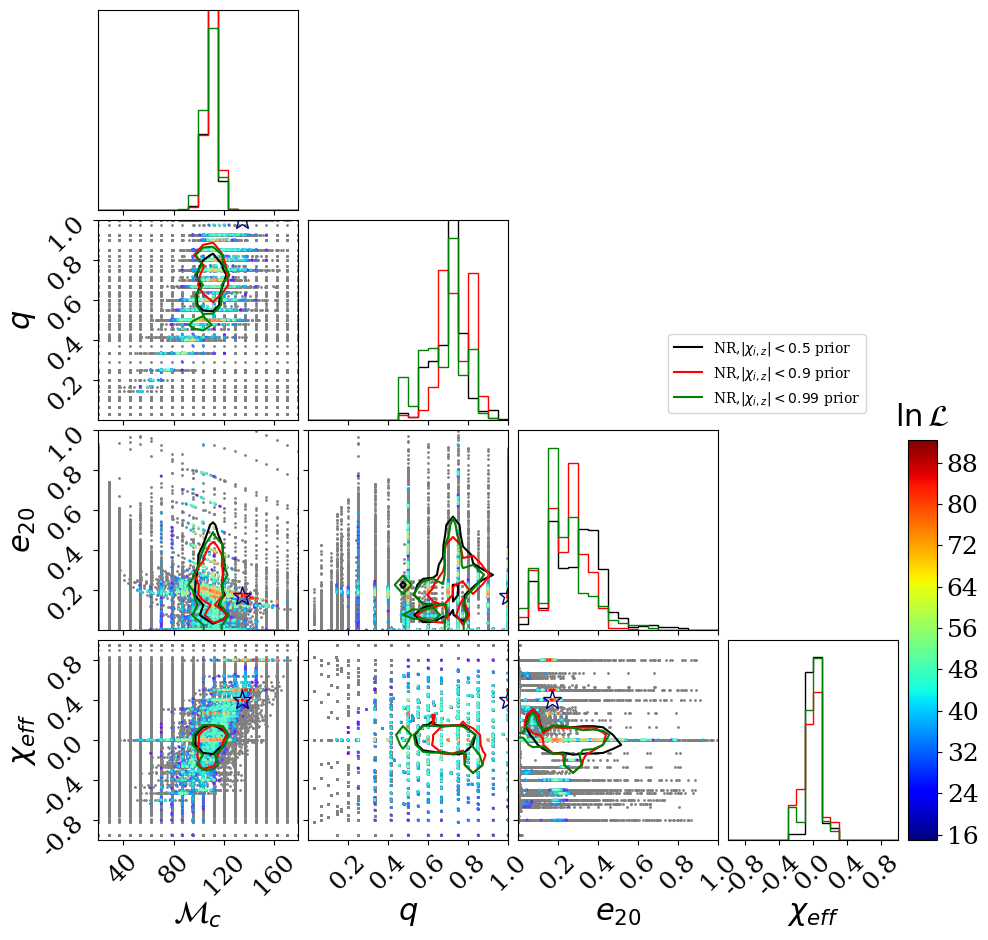}
    \includegraphics[width=1.45\columnwidth]{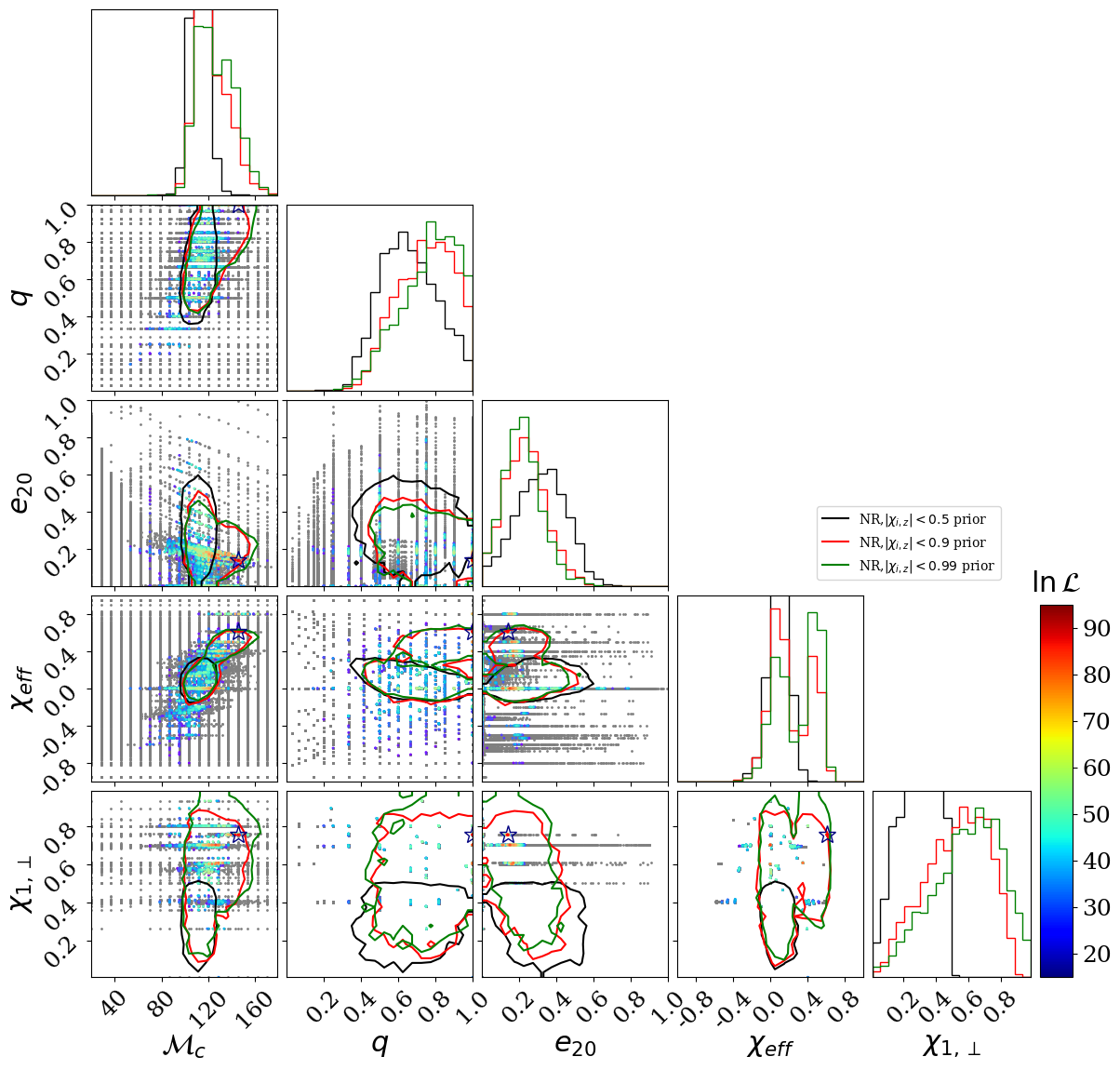}
    \caption{A corner plot showing the results of NR based PE for 
    GW190521 with a grid consisting of the RIT catalog \cite{Ficarra:2026cej} with the 
    restrictions on mass ratio from \secref{ss:analysis} for a total of 
    1519 simulations for the nonprecessing analysis ({\it top}) 
    and 2117 simulations for the precessing spin 
    analysis ({\it bottom}). These are rescaled over the detector frame chirp mass 
    range of $\mc \in [20,178.7] \ M_\odot$. Three NR posteriors are shown with 
    different $|\chi_{i,z}|$ priors, $|\chi_{i,z}|< 0.99$ (green), $|\chi_{i,z}|< 0.9$ 
    (red), and $|\chi_{i,z}|< 0.5$ (black). The colorbar represents the value of 
    $\lnLmarg$ and a star is placed at the peak $\lnL$ simulation, these are present 
    on all corner plots.}
    \label{fig:190521_PE}
\end{figure*}

\Figref{fig:190521_PE} shows our analyses of GW190521 with direct comparison to our 
bank of nonprecessing simulations in the top panel and the set of nonprecessing plus 
precessing simulations in the bottom panel, see \secref{ss:analysis} for details on 
grid construction. Both analyses are performed on a grid covering the detector frame 
chirp mass range of $\mc \in [20,178.7]\ M_\odot$ which is consistent with the LVK 
analysis. We first focus on the nonprecessing analysis where the marginal likelihoods
derived by direct comparison of each NR simulation to the data reveal the best
fitting simulations for this analysis are mildly eccentric at 20 Hz, with 
$e_{20} \in (0.14,0.20)$, many have positive effective spin, and the top simulations 
have equal mass ratios $q \approx 1$ with a few at $q=0.5$. These observations are 
seen by the 10 peak likelihood simulations along with their exact intrinsic parameters 
in \Tabref{tab:190521_aligned_peakLnL}. The best matching nonprecessing simulation  
with a marginal likelihood of $\lnL =92.360$ is RIT:BBH:1794 which has equal masses, 
effective spin of $\chieff=0.5$, and an eccentricity at 20 Hz of $e_{20}=0.161$.

\begin{table*}
  \caption{Peak marginal likelihood, $\lnLmarg$, NR simulations 
  of GW190521 for the $\fref=20$ Hz nonprecessing analysis shown in the top panel of
  \figref{fig:190521_PE}. Where all have the dimensionless spin components 
  $s_1^x=s_1^y=s_2^x=s_2^y=0$. The likelihood values reported in the  
  $\lnL$ column are a result of Gaussian process fitting discussed in 
  \secref{s:events}. $\frefnr$ is the initial apoapsis NR frequency, which is used in the conversion of eccentricity to an apoapsis frequency of $\fref=20$ Hz. 
  $e_{20}$ is the conversion of the initial 3.5PN order eccentricity, 
  $e_{3.5\rm{PN}}$, to 20 Hz using \eqnref{eq:e20}.}
  \label{tab:190521_aligned_peakLnL} 
\begin{ruledtabular}
\begin{tabular}{cccccccccccc}

NR\_sim\_id	&	$m_1/M_\odot$&	$m_2/M_\odot$&	$q$&	$s_1^z$&	$s_2^z$&	$e_{3.5\rm{PN}}$&  $e_{20}$& $\lnL$& $\frefnr$ (Hz)\\
\hline

RIT:BBH:1794& 161.924& 161.924& 1.000& 0.500& 0.500& 0.803& 0.161& 92.360& 1.79\\
RIT:BBH:1833& 154.403& 154.317& 0.999& 0& 0.801& 0.802& 0.169& 92.358& 1.88\\
RIT:BBH:1817& 176.490& 176.490& 1.000& 0.801& 0.801& 0.835& 0.147& 92.001& 1.53\\
RIT:BBH:1795& 154.369& 154.369& 1.000& 0.500& 0.500& 0.839& 0.180& 91.808& 1.87\\
RIT:BBH:1818& 163.971& 163.971& 1.000& 0.801& 0.801& 0.863& 0.161& 91.385& 1.59\\
RIT:BBH:1723& 178.829& 89.367& 0.500& 0& 0.801& 0.792& 0.199& 90.753& 2.26\\
RIT:BBH:1834& 147.747& 147.665& 0.999& 0& 0.801& 0.834& 0.186& 90.700& 1.94\\
RIT:BBH:1722& 190.238& 95.121& 0.500& 0.400& 0& 0.717& 0.160& 90.393& 2.13\\
RIT:BBH:1725& 183.902& 91.952& 0.500& 0.400& -0.400& 0.718& 0.166& 90.121& 2.20\\
RIT:BBH:1383& 157.469& 110.229& 0.700& 0& 0& 0.759& 0.186& 89.775& 2.26\\

\end{tabular}
\end{ruledtabular}
\end{table*}

\begin{figure}[h!]
    \includegraphics[width=0.46\textwidth]{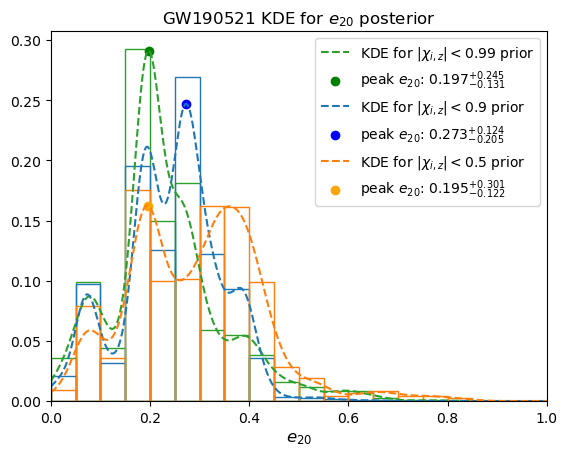}
    \includegraphics[width=0.46\textwidth]{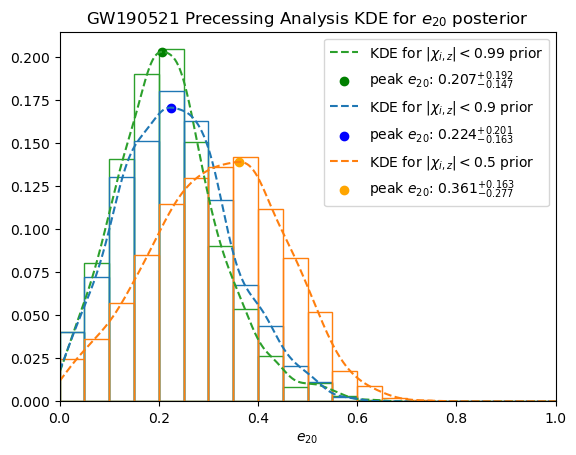}
    \caption{Kernel density estimate of the GW190521 NR-based PE 
    1-dimensional eccentricity posterior at 20\,Hz from the nonprecessing analysis 
    ({\it top}) and precessing spin analysis ({\it bottom}) 
    (histograms from \figref{fig:190521_PE}) for the three different spin priors, 
    $|\chi_{i,z}|< 0.99$ (green), $|\chi_{i,z}|< 0.9$ (blue) and $|\chi_{i,z}|< 0.5$
    (orange). The maximum of the KDE for each prior is indicated by a dot of the same 
    color with errors reported as the 90\% confidence interval of the histogram.}
    \label{fig:190521_e20_kde}
\end{figure}

We now look at the inferred joint posteriors generated by interpolating the marginal 
likelihoods over mass, aligned spins, and eccentricity. Noting that our NR grid may 
potentially introduce systematic errors due to gaps in coverage of parameter space, we 
estimate that the binary mass ratio posterior favors mildly asymmetric components with 
agreement across the spin priors $|\chi_{i,z}|<0.5$, $<0.9$, and $<0.99$, 
$q=0.749_{-0.155}^{+0.085}$, $q=0.750_{-0.140}^{+0.100}$, and 
$q=0.747_{-0.248}^{+0.099}$, respectively. Similarly, the effective spin posterior 
finds good agreement between the priors and is centered at zero-effective spin with 
$\chieff =0.001_{-0.030}^{+0.175}$ for $|\chi_{i,z}|<0.5$, 
$\chieff =0.003_{-0.214}^{+0.194}$ for $|\chi_{i,z}|<0.9$,
and $\chieff =0.003_{-0.212}^{+0.145}$ for $|\chi_{i,z}|<0.99$. The choice of spin 
prior has minor effects on the value of chirp mass, with 
$\mc=111.345_{-8.434}^{+4.780}\ M_\odot$, $\mc=109.756_{-6.742}^{+7.236}\ M_\odot$, 
and $\mc=108.644_{-9.011}^{+7.966}\ M_\odot$ respectively for each of the spin priors 
$|\chi_{i,z}|<0.5$, $<0.9$, and $<0.99$. While finally the value of eccentricity 
at 20 Hz is clearly non-zero where $e_{20}=0.195_{-0.122}^{+0.301}$, 
$e_{20}=0.273_{-0.205}^{+0.124}$, and $e_{20}=0.197_{-0.131}^{+0.245}$ 
respectively for each of the spin priors $|\chi_{i,z}|<0.5$, $<0.9$, and $<0.99$. 
We note here that the higher value of eccentricity found for the 
$|\chi_{i,z}|<0.9$ prior may be due to the exclusion of simulations with 
high spins that lie above the prior range. While the posterior generated with 
$|\chi_{i,z}|<0.5$ agrees with our least restrictive posterior ($|\chi_{i,z}|<0.99$) 
as there is a secondary probability density peak at $e_{20}=0.36$ where the difference
in probability density between the primary and secondary peak is 0.0005 
(see top panel of \figref{fig:190521_e20_kde}) indicating near 
equal support for the two eccentricity values. The extended support for 
eccentricity in the posterior generated with $|\chi_{i,z}|<0.5$ is corroborated 
by the Bayes factors calculated using \Eqnref{eq:bayes} where $\bayesecc= 0.64$, 
$\bayesecc= 0.37$, and $\bayesecc= 0.26$ respectively for each of the spin priors 
$|\chi_{i,z}|<0.5$, $<0.9$, and $<0.99$. 
The Bayes factors found for each spin prior 
are positive indicating support for eccentricity in this non-precessing analysis. We 
provide in \Tabref{tab:190521_kde_params} a summary of the maximum posterior values 
found through a KDE analysis of the posteriors in the nonprecessing analysis from the 
top panel of \figref{fig:190521_PE} and the associated Bayes factors found through 
\Eqnref{eq:bayes}.

\begin{table}[h]
\renewcommand{\arraystretch}{1.5}
  \caption{Summary of Maximum Kernel Density Estimates for the nonprecessing analysis of GW190521 shown in the top panel of \figref{fig:190521_PE} and Bayes factors for the quasi-circular vs eccentric hypothesis calculated from \Eqnref{eq:bayes}. The $e_{20}$ KDE is shown in the top panel of \figref{fig:190521_e20_kde}.}
  \label{tab:190521_kde_params} 
\begin{ruledtabular}
\begin{tabular}{c|ccc}
Parameters & $|\chi_{i,z}|< 0.5$ & $|\chi_{i,z}|< 0.9$ & $|\chi_{i,z}|< 0.99$ \\  
  & posterior & posterior & posterior \\
\hline
$\mc (M_\odot)$ & $111.345_{-8.434}^{+4.780}$ & $109.756_{-6.742}^{+7.236}$ & $108.644_{-9.011}^{+7.966}$ \\
$q$ & $0.749_{-0.155}^{+0.085}$ & $0.750_{-0.140}^{+0.100}$ &  $0.747_{-0.248}^{+0.099}$ \\
$e_{20}$ & $0.195_{-0.122}^{+0.301}$ & $0.273_{-0.205}^{+0.124}$ & $0.197_{-0.131}^{+0.245}$ \\
$\chieff$ & $0.001_{-0.030}^{+0.175}$ & $0.003_{-0.214}^{+0.194}$ & $0.003_{-0.212}^{+0.145}$ \\
$\bayesecc$ & 0.64 & 0.37 & 0.26 \\
\end{tabular}
\end{ruledtabular}
\end{table}

Now we look at the precessing analysis presented in the bottom panel of 
\figref{fig:190521_PE} where we add our bank of precessing simulations to the grid 
used in the nonprecessing analysis. The marginal likelihoods found through direct 
comparison of each NR simulation in the precessing analysis show the best fitting 
simulations of this analysis to be mildly eccentric at 20 Hz, with 
$e_{20} \in (0.11,0.17)$, many having positive effective spin, large 
$\chi_{1,\perp}$, and the top ten best fitting simulations have equal mass 
black holes, as seen in \Tabref{tab:190521_prec_peakLnL}. Reaching a higher 
likelihood than the best matching nonprecessing simulation is RIT:BBH:1618---the 
best matching precessing simulation---with a likelihood of $\lnL=95.1$, equal 
masses, effective spin of $\chieff = 0.61$, in-plane spin of the primary mass
$\chi_{1,\perp}=0.75$, and eccentricity at 20 Hz of $e_{20}=0.139$.
We note here that the top $\lnL$ NR simulations RIT:BBH:1618 and RIT:BBH:1616 in 
\Tabref{tab:190521_prec_peakLnL} both correspond to a primary black hole with 
intrinsic spins $s_1=0.97$ and secondary $s_2=0.97$ in a highly precessing eccentric 
configuration. 

\begin{table*} 
  \caption{Peak marginal likelihood, $\lnLmarg$, NR simulations 
  of GW190521 for the $\fref=20$ Hz precessing analysis shown in the bottom panel of
  \figref{fig:190521_PE}. The likelihood values reported in the  
  $\lnL$ column are a result of Gaussian process fitting discussed in 
  \secref{s:events}. $\frefnr$ is the initial apoapsis NR frequency, which is used in the conversion of eccentricity to an apoapsis frequency of $\fref=20$ Hz. 
  $e_{20}$ is the conversion of the initial 3.5PN order eccentricity, 
  $e_{3.5\rm{PN}}$, to 20 Hz using \eqnref{eq:e20}.}
  \label{tab:190521_prec_peakLnL} 
\begin{ruledtabular}
\begin{tabular}{cccccccccccccccc}

NR\_sim\_id	& $m_1/M_\odot$& $m_2/M_\odot$&	$q$& $s_1^x$& $s_1^y$& $s_1^z$& $s_2^x$& $s_2^y$& $s_2^z$& $e_{3.5\rm{PN}}$& $e_{20}$& $\lnL$& $\frefnr$\\
 & & & & & & & & & & & & & (Hz)\\
\hline
RIT:BBH:1618& 164.088& 164.088& 1.000& -0.754& -0.00031& 0.610& 0.754& 0.00031& 0.610& 0.750& 0.139& 95.100& 1.73\\
RIT:BBH:1616& 166.896& 166.896& 1.000& 0.754& 0.00031& 0.610& -0.754& -0.00031& 0.610& 0.750& 0.139& 95.084& 1.72\\
RIT:BBH:1637& 145.014& 145.014& 1.000& 0.700& 0.001& -0.00006& 0.700& 0.001& -0.00006& 0.765& 0.151& 93.450& 1.82\\
RIT:BBH:1648& 141.431& 141.431& 1.000& 0.607& 0.350& 0& 0.700& 0& 0& 0.640& 0.115& 93.225& 1.83\\
RIT:BBH:1627& 141.425& 141.425& 1.000& 0.600& 0.001& -0.00004& 0.600& 0.001& -0.00004& 0.766& 0.164& 92.976& 1.96\\
RIT:BBH:1603& 147.706& 147.706& 1.000& 0.001& -0.607& 0.350& -0.001& 0.607& 0.350& 0.763& 0.169& 92.789& 2.03\\
RIT:BBH:1659& 143.261& 143.260& 0.999& 0.350& 0.607& 0& 0.700& 0& 0& 0.767& 0.157& 92.734& 1.88\\
RIT:BBH:1794& 161.879& 161.879& 1.000& 0& 0& 0.500& 0& 0& 0.500& 0.803& 0.161& 92.370& 1.79\\
RIT:BBH:1620& 163.779& 163.779& 1.000& -0.001& 0.754& 0.611& 0.001& -0.754& 0.611& 0.750& 0.137& 92.370& 1.69\\
RIT:BBH:1833& 154.427& 154.340& 0.999& 0& 0& 0& 0& 0& 0.801& 0.802& 0.169& 92.351& 1.88\\

\end{tabular}
\end{ruledtabular}
\end{table*}

We infer a joint posterior distribution on our parameters by interpolating these 
sparsely sampled marginal likelihoods over mass, aligned spins, eccentricity, 
and the spin precession parameter $\chi_{1,\perp}$. The mass ratio 
posterior agrees well with the nonprecessing analysis, returning mass ratios 
$q=0.620_{-0.198}^{+0.297}$, $q=0.817_{-0.324}^{+0.145}$, and 
$q=0.808_{-0.324}^{+0.163}$ respectively for $|\chi_{i,z}|<0.5$, $<0.9$,
$<0.99$. The precessing analysis finds more support for positive effective spin where 
$\chieff=0.067_{-0.161}^{+0.192}$, $\chieff=0.081_{-0.162}^{+0.477}$, and 
$\chieff=0.491_{-0.549}^{+0.082}$ respectively for $|\chi_{i,z}|<0.5$, $<0.9$, and 
$<0.99$. The chirp mass values lie in the same region as the 
nonprecessing analysis but have a wider 90\% confidence interval, with 
$\mc=111.345_{-10.567}^{+11.692}\ M_\odot$, $\mc=113.092_{-8.188}^{+35.496}\ M_\odot$, 
and $\mc=115.316_{-10.660}^{+37.497}\ M_\odot$ respectively for $|\chi_{i,z}|<0.5$, 
$<0.9$, and $<0.99$. The $\chi_{1,\perp}$ posterior favors large transverse spin with 
$\chi_{1,\perp}=0.293_{-0.198}^{+0.173}$, $\chi_{1,\perp}=0.651_{-0.465}^{+0.157}$, 
and $\chi_{1,\perp}=0.678_{-0.462}^{+0.218}$ for $|\chi_{i,z}|<0.5$, $<0.9$, 
and $<0.99$, respectively, where the systematic decrease in parameter value is 
due to the successive exclusion of high spins with each spin prior. 
The precessing Bayes factors found for each choice of spin prior are all positive 
confirming our observations on the transverse spin posteriors,
where $\bayesprec=3.81$ for both $|\chi_{i,z}|<0.5$ and $|\chi_{i,z}|<0.9$, and 
$\bayesprec=3.72$ for $|\chi_{i,z}|<0.99$.
Finally the eccentricity posterior favors a higher eccentricity than found in the 
nonprecessing analysis with values of $e_{20}=0.361_{-0.277}^{+0.163}$, 
$e_{20}=0.224_{-0.163}^{+0.201}$, and $e_{20}=0.207_{-0.147}^{+0.192}$ for 
$|\chi_{i,z}|<0.5$, $<0.9$, and $<0.99$, respectively. 
We note here that the peak value of eccentricity decreases with an increase in the 
value of spin prior, likely due to our best fitting simulations for this event (as 
shown in \Tabref{tab:190521_prec_peakLnL}) having large values of $|\chi_{i,z}|$. 
This behavior is seen best in the eccentricity KDE for the precessing analysis 
in the bottom panel of \figref{fig:190521_e20_kde}. Again the eccentric Bayes factors 
found for each spin prior confirm our observations on the eccentricity posteriors, 
where $\bayesecc=0.38$, $\bayesecc=0.21$, and $\bayesecc=0.24$ for $|\chi_{i,z}|<0.5$, 
$<0.9$, and $<0.99$, respectively. We find all three spin prior choices return a 
positive eccentric Bayes factor and a positive precessing Bayes factor indicating this 
event has support for eccentricity and precession.
We note that the posterior generated with the $|\chi_{i,z}|<0.99$ prior in this 
precessing analysis allows for the inclusion of the most simulations and choose this 
posterior as our preferred posterior. A summary of our posterior peaks for the 
precessing analysis are shown in \Tabref{tab:190521_prec_kde_params} along with the 
Bayes factors for the quasicircular vs eccentric hypothesis and nonprecessing vs 
precessing hypothesis.

\begin{table}[h]
\renewcommand{\arraystretch}{1.5}
  \caption{Summary of Maximum Kernel Density Estimates for the precessing analysis 
  of GW190521 shown in the bottom panel of \figref{fig:190521_PE}. Bayes factors for the quasi-circular vs eccentric hypothesis, $\bayesecc$, calculated from \Eqnref{eq:bayes} and the nonprecessing vs precessing hypothesis, $\bayesprec$, calculated from \eqnref{eq:bayesprec}. The $e_{20}$ KDE is shown in the bottom panel of \figref{fig:190521_e20_kde}.}
  \label{tab:190521_prec_kde_params} 
\begin{ruledtabular}
\begin{tabular}{c|ccc}
Parameters & $|\chi_{i,z}|< 0.5$ & $|\chi_{i,z}|< 0.9$ & $|\chi_{i,z}|< 0.99$ \\  
  & posterior & posterior & posterior \\
\hline
$\mc (M_\odot)$ & $111.345_{-10.567}^{+11.692}$ & $113.092_{-8.188}^{+35.496}$ & $115.316_{-10.660}^{+37.497}$ \\
$q$ & $0.620_{-0.198}^{+0.297}$ & $0.817_{-0.324}^{+0.145}$ & $0.808_{-0.324}^{+0.163}$ \\
$e_{20}$ & $0.361_{-0.277}^{+0.163}$ & $0.224_{-0.163}^{+0.201}$ & $0.207_{-0.147}^{+0.192}$ \\
$\chieff$ & $0.067_{-0.161}^{+0.192}$ & $0.081_{-0.162}^{+0.477}$ & $0.491_{-0.549}^{+0.082}$ \\
$\chi_{1, \perp}$& $0.293_{-0.198}^{+0.173}$ & $0.651_{-0.465}^{+0.157}$ & $0.678_{-0.462}^{+0.218}$ \\
$\bayesecc$ & 0.38 & 0.21 & 0.24\\
$\bayesprec$ & 3.81 & 3.81 & 3.72\\
\end{tabular}
\end{ruledtabular}
\end{table}

After the analysis of GW190521 in both a nonprecessing and precessing regime, 
we find the precessing analysis returns simulations with higher likelihoods 
thus this event is better matched to an eccentric precessing waveform.
Particularly we find that including aligned spins up to their maximal value of 0.99 
in the choice of spin prior is vital to correctly estimating the parameters, 
especially the transverse spin $\chi_{1,\perp}$ and eccentricity at 20 Hz $e_{20}$.
This choice in preferred spin prior is motivated by the need to include the maximal 
amount of simulations when generating a posterior, using the least restrictive spin 
prior of $|\chi_{i,z}|<0.99$ allows for this.

We can compare our preferred results from the precessing analysis using the 
$|\chi_{i,z}|<0.99$ spin prior to the results from the LVK
\cite{LIGO-O3-GW190521-implications} noting that eccentricity is not included in their 
analysis. We find a very similar mass ratio, and chirp mass, where our (LVK) median 
values are $q=0.808$ ($q=0.79$), and $\mc=115.316 \ M_\odot$ ($\mc=117\ M_\odot$). 
We find a higher value of effective spin where our (LVK) median value is 
$\chieff=0.491$ ($\chieff=0.008$). We find a median primary mass in-plane spin of
$\chi_{1,\perp}=0.678$, while the LVK \cite{LIGO-O3-GW190521-implications} does not 
provide $\chi_{1,\perp}$ directly the value can be calculated from the in-plane spin 
posteriors of the primary mass provided in the data release 
\cite{GW190521_NRSur7dq4_PE} they find $\chi_{1,\perp}=0.55$ noting that 
$\chi_{\rm p}=0.68$. Similarly we compare our results from the preferred posterior 
to Gayathri {\it et al.} \cite{Gayathri:2020coq}, they find median values of $q=1$, 
$\mc=144\ M_\odot$\footnote{The detector frame chirp mass was 
calculated with the inferred mass ratio, source frame total mass, and redshift 
provided in Table 1 of \cite{Gayathri:2020coq}}, $\chieff=0.27$, $\chi_{\rm p}=0.66$, 
and eccentricity $e=0.69$. Ref~\cite{Gayathri:2020coq} finds higher chirp mass and 
mass ratio, lower effective spin, and nearly equivalent effective spin precession 
$\chi_{\rm p}$ (as they find nearly the same value of $\chi_{\rm p}$ as the LVK 
\cite{LIGO-O3-GW190521-implications}), our eccentricity is not directly comparable to 
\cite{Gayathri:2020coq} as they calculate $e$ at 10 Hz assuming a total mass of 
$50\ M_\odot$. 

\begin{figure*}[h!tbp]
    \includegraphics[width=\textwidth]{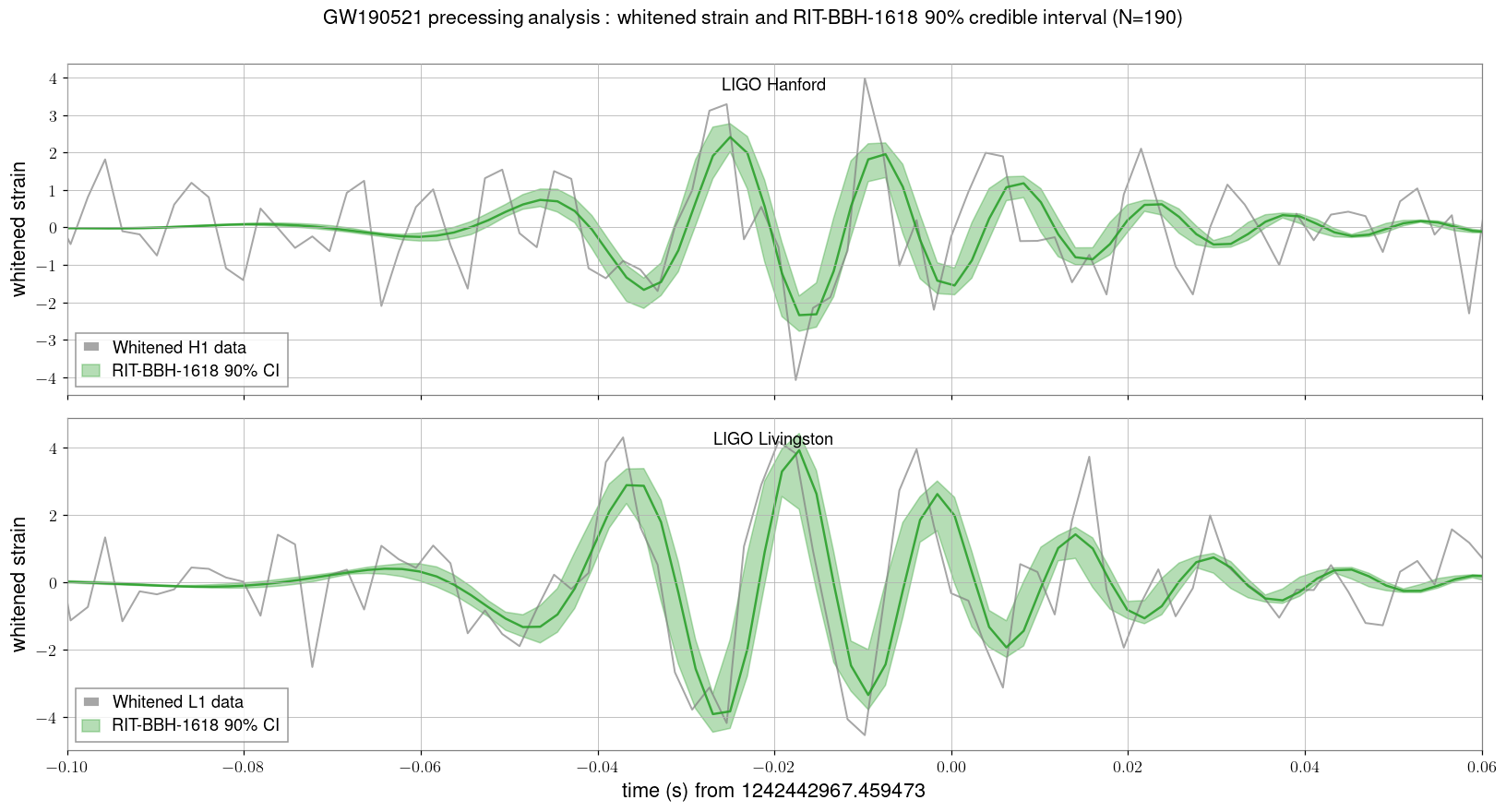}
    \caption{Best fitting time-domain NR waveform RIT:BBH:1618 median (dark green) and 90\% credible interval (light green band) overlaid on whitened strain data (grey) near GW190521.}
    \label{fig:190521_prec_wf}
\end{figure*}

Once our full precessing PE analysis is complete a grid of the peak likelihood 
simulation alone is used in an additional ILE run that marginalizes the coherent 
multi-detector likelihood over the extrinsic parameters to return posterior samples 
including both the extrinsic and intrinsic parameters of the peak likelihood 
simulation. Importantly, marginalization over coalescence time requires resampling to 
generate one phase-coherent time per posterior sample and preserve the correct 
absolute normalization of the likelihood otherwise each sample would need to have a 
hand-picked common reference time which desynchronizes phase across posterior draws. 
Since we run ILE with a single grid point a limited number of fair-draw samples are 
returned, so we must loop over several independent ILE calls until a target number of 
samples are reached, in this work we require at least $N=150$ samples to be generated. 
These posteriors are then used to generate the peak NR waveform strain $h(t)$ with 
the extrinsic parameters of each sample, where each waveform is Fourier transformed 
to the frequency domain, whitened with the same PSD used in the PE analysis then 
band-limited to the low and high frequency cutoffs set in \secref{ss:analysis} and 
transformed back to the time-domain. Each waveform realization is interpolated onto 
a common time grid where at each time sample the $5^{\rm th}$ and $95^{\rm th}$ 
percentiles are calculated to form the 90\% credible interval. Similarly the LIGO 
strain data is transformed to the frequency domain, whitened with the analysis PSDs, 
band-limited to match analysis settings, and transformed back to the time-domain. 
This process is used to generate the reconstructed waveforms shown in Figures 
\ref{fig:190521_prec_wf} and \ref{fig:231123_e20_prec_wf}.

We present in \figref{fig:190521_prec_wf} a waveform reconstruction of our best 
matching simulation by likelihood, RIT:BBH:1618, along with the band-limited whitened 
strain data near the GW190521 detector response in the LIGO Hanford (H1) and LIGO 
Livingston (L1) instruments. We achieved 190 fair-draw samples for our peak likelihood 
eccentric precessing simulation, which has $\lnL=95.1$, $q=1$, $\chieff=0.61$, 
$\chi_{1,\perp}=0.75$, and $e_{20}=0.139$. Our reconstruction fits the LIGO strain 
data well in both amplitude and phase, suggesting this simulation is truly a good 
match to the data.

\subsection{GW231123}
\label{ss:gw231123}
The LVK collaboration identified in GWTC-4.0 \cite{LIGO-O4-O4a_catalog} 
GW231123 an event with SNR $20.7_{-0.3}^{+0.2}$ where the component 
masses, mass ratio and chirp mass in the detector frame as reported in 
\cite{O4a_PE_data_release} are: $m_1=180.9_{-20.8}^{+31.5}\ M_\odot$, 
$m_2=146.2_{-50.1}^{+24.3}\ M_\odot$, $q=0.79_{-0.25}^{+0.18}$, and 
$\mc=141.3_{-29.1}^{+17.5}\ M_\odot$. This event also has significant support for 
positive effective spin with an 88\% probability that $\chieff > 0$ where the 
reported effective spin from Table 3 of \cite{LIGO-O4-O4a_catalog} is 
$\chieff=0.31_{-0.41}^{+0.25}$. GW231123 was further analyzed in 
\cite{GW231123_special}, they find detector frame component masses, mass ratio, 
and chirp mass of $m_1=186_{-24}^{+49}\ M_\odot$, $m_2=139_{-59}^{+30}\ M_\odot$, 
$q=0.74_{-0.38}^{+0.22}$, and $\mc=137_{-28}^{+21}\ M_\odot$. They find consistent 
support for high spins in both component masses regardless of the model used with a 
91\% probability of $\chi_1\geq0.7$ and a 63\% probability of $\chi_2\geq0.7$ with 
an 89\% probability of $\chieff>0$ where $\chieff=0.32_{-0.41}^{+0.25}$. Their 
effective precessing spin is $\chi_{\rm p}=0.77_{-0.19}^{+0.18}$. The analysis in 
\cite{GW231123_special} conclude that this event has very high component spins with
some support for precession that can be more confidently claimed with further study.

Here we discuss the analyses conducted using a low frequency cutoff $\flow$ and 
reference frequency $\fref$ of 20 Hz, the analyses using a 10 Hz $\flow$ and $\fref$ 
are presented and discussed in \appendref{s:231123_e10}.

\begin{figure*}[h!tbp]
    \includegraphics[width=1.45\columnwidth]{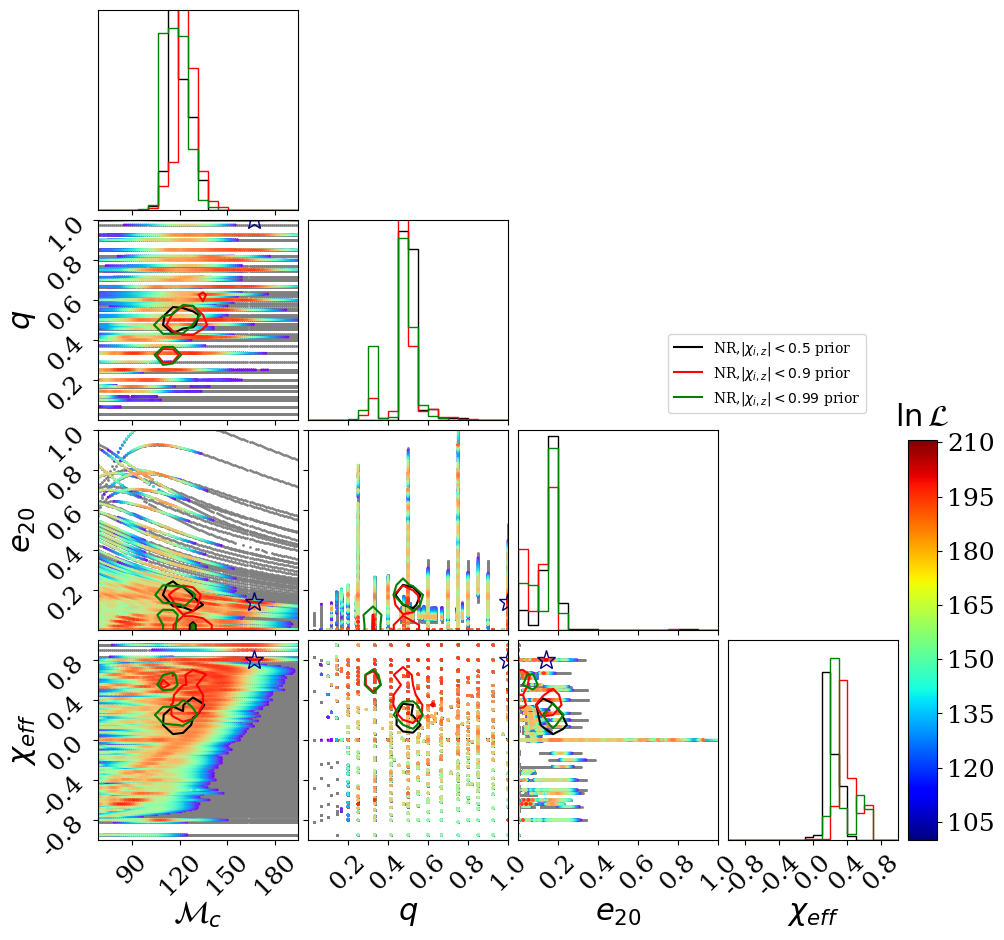}
    \includegraphics[width=1.45\columnwidth]{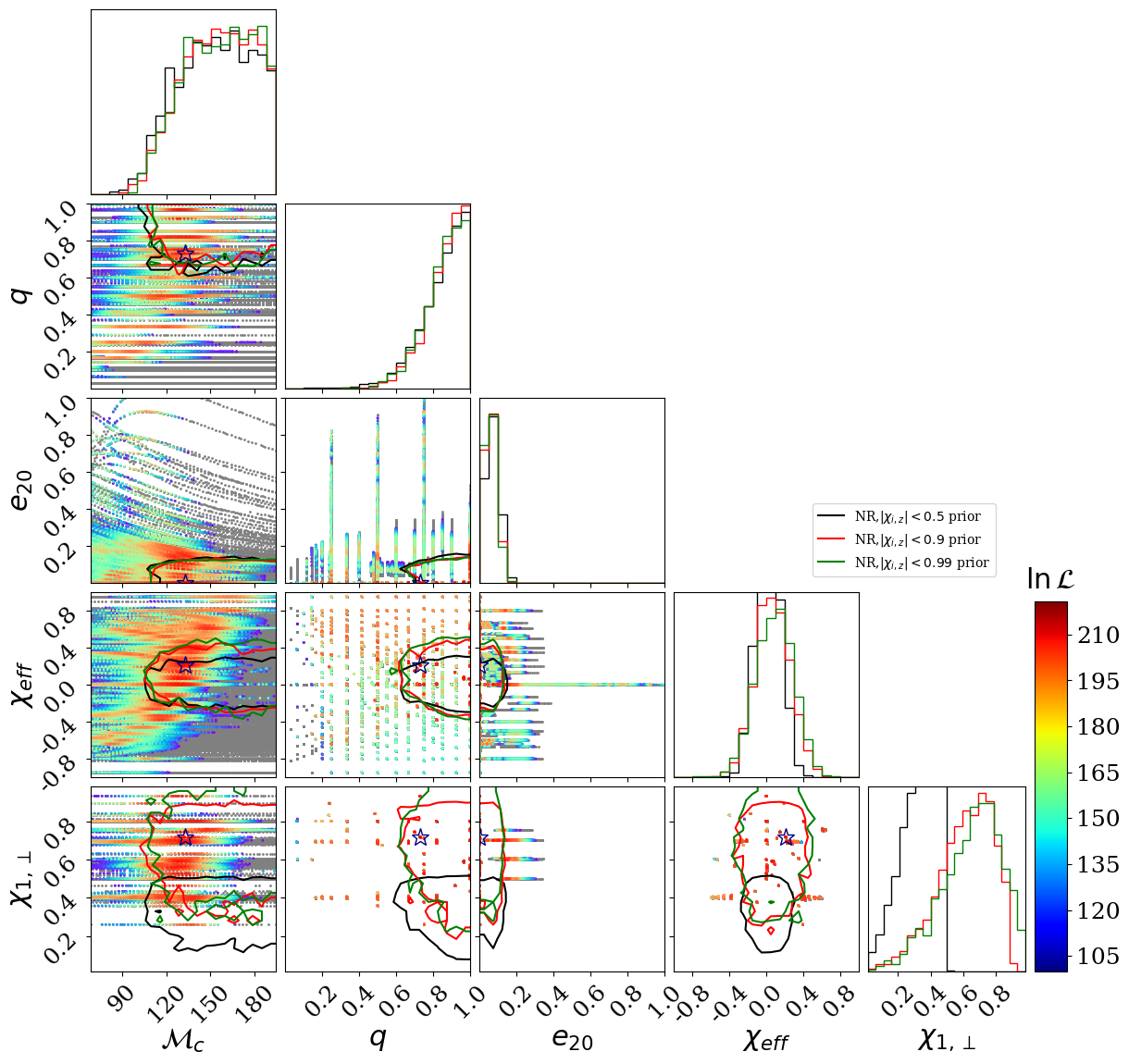}
    \caption{A corner plot showing the results of NR based PE for 
    GW231123 with a grid consisting of the RIT catalog \cite{Ficarra:2026cej} 
    with the restrictions on mass ratio from \secref{ss:analysis} for a total 
    of 1519 simulations for the nonprecessing analysis at $\fref=20$ Hz 
    ({\it top}) and 2117 simulations for the precessing spin analysis 
    at $\fref=20$ Hz ({\it bottom}). These are rescaled over the detector 
    frame chirp mass range of $\mc \in [68.2,194.6] \ M_\odot$. Three NR posteriors 
    are shown with different $|\chi_{i,z}|$ priors, $|\chi_{i,z}|< 0.99$ (green), 
    $|\chi_{i,z}|< 0.9$ (red), and $|\chi_{i,z}|< 0.5$ (black).}
    \label{fig:231123_e20_PE}
\end{figure*}

\begin{table*}[hbtp] 
  \caption{Peak marginal likelihood, $\lnLmarg$, NR simulations 
  of GW231123 for the $\fref=20$ Hz nonprecessing analysis shown in the top panel of
  \figref{fig:231123_e20_PE}. Where all have the dimensionless spin components 
  $s_1^x=s_1^y=s_2^x=s_2^y=0$. The likelihood values reported in the  
  $\lnL$ column are a result of Gaussian process fitting discussed in 
  \secref{s:events}. $\frefnr$ is the initial apoapsis NR frequency, which is used in the conversion of eccentricity to an apoapsis frequency of $\fref=20$ Hz. 
  $e_{20}$ is the conversion of the initial 3.5PN order eccentricity, 
  $e_{3.5\rm{PN}}$, to 20 Hz using \eqnref{eq:e20}.}
  \label{tab:231123_e20_aligned_peakLnL} 
\begin{ruledtabular}
\begin{tabular}{cccccccccccccc}

NR\_sim\_id	&	$m_1/M_\odot$&	$m_2/M_\odot$&	$q$&	$s_1^z$&	$s_2^z$&	$e_{3.5\rm{PN}}$& $e_{20}$& $\lnL$& $\frefnr$ (Hz)\\
\hline

RIT:BBH:1816& 191.306& 191.306& 1.000& 0.801& 0.801& 0.803& 0.140& 210.591& 1.56\\
RIT:BBH:1859& 197.881& 49.499& 0.250& -0.801& 0.000& 0.915& 0.042& 205.884& 0.39\\
RIT:BBH:1724& 212.904& 106.454& 0.500& 0.801& -0.801& 0.714& 0.142& 205.335& 1.90\\
RIT:BBH:0796& 235.865& 118.046& 0.500& 0.952& 0.000& 0.001& 0.000& 204.869& 5.33\\
RIT:BBH:1721& 196.697& 98.349& 0.500& 0.400& 0.400& 0.717& 0.152& 204.722& 2.02\\
RIT:BBH:1880& 185.427& 61.845& 0.334& -0.801& 0.000& 0.916& 0.053& 204.454& 0.49\\
RIT:BBH:1915& 262.547& 87.554& 0.333& 0.950& -0.950& 0.000& 0.000& 204.326& 5.36\\
RIT:BBH:1817& 186.907& 186.907& 1.000& 0.801& 0.801& 0.835& 0.138& 203.929& 1.45\\
RIT:BBH:0314& 233.794& 116.985& 0.500& 0.851& 0.500& 0.001& 0.000& 203.802& 3.64\\
RIT:BBH:1008& 228.323& 125.580& 0.550& 0.800& 0.800& 0.001& 0.000& 203.588& 4.46\\

\end{tabular}
\end{ruledtabular}
\end{table*}

\Figref{fig:231123_e20_PE} shows our analyses of GW231123 with direct comparison to 
our bank of nonprecessing simulations in the top panel and the set of nonprecessing 
plus precessing simulations in the bottom panel, see \secref{ss:analysis} for details 
on grid construction. Both analyses are performed on a grid covering the detector 
frame chirp mass range of $\mc \in [68.2,194.6]\ M_\odot$ which is consistent with 
the LVK analyses. We first focus on the nonprecessing analysis, direct comparison of 
each NR simulation to the data return marginal likelihoods which show that the best 
fitting simulations in the nonprecessing analysis range from mildly eccentric to 
noneccentric at 20 Hz, with $e_{20}\in (0,0.16)$, many have large effective spin 
with a few having large negative effective spin, mass ratios of the top simulations 
are split equally across equal mass, $q\approx0.5$, and $q\approx 0.3$. A summary of 
these observations are provided by the 10 peak likelihood simulations along with 
their exact intrinsic parameters in \Tabref{tab:231123_e20_aligned_peakLnL}. The 
peak likelihood nonprecessing simulation (RIT:BBH:1816) has marginal likelihood of 
$\lnL=210.591$, outperforming all other nonprecessing simulations by a $\lnL$ 
value nearly 5 above the rest. RIT:BBH:1816 has equal masses, effective spin of 
$\chieff=0.801$, and eccentricity at 20 Hz of $e_{20}=0.140$. 

\begin{figure}[h!tbp]
    \includegraphics[width=0.46\textwidth]{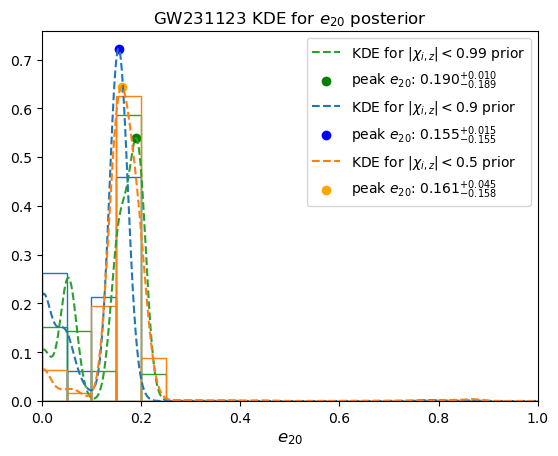}
    \includegraphics[width=0.46\textwidth]{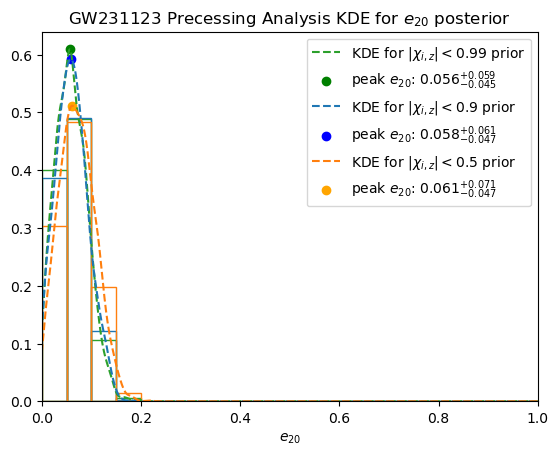}
    \caption{Kernel density estimate of the GW231123 NR-based PE 
    1-dimensional eccentricity posterior at 20\,Hz from the nonprecessing analysis 
    ({\it top}) and the precessing spin analysis ({\it bottom})
    (histograms from \figref{fig:231123_e20_PE}) for the three different 
    spin priors, $|\chi_{i,z}|< 0.99$ (green), $|\chi_{i,z}|< 0.9$ (blue) and 
    $|\chi_{i,z}|< 0.5$ (orange). The maximum of the KDE for each prior is indicated 
    by a dot of the same color with errors reported as the 90\% confidence interval of the histogram.}
    \label{fig:231123_e20_kde}
\end{figure}

Interpolating the sparsely sampled marginal likelihoods versus mass, aligned spins, 
and eccentricity, we infer a joint posterior distribution over these parameters.
We find agreement for mass ratio across the three spin priors used to construct our 
posteriors with $q=0.499_{-0.010}^{+0.084}$ for $|\chi_{i,z}|< 0.5$, 
$q=0.496_{-0.166}^{+0.102}$ for $|\chi_{i,z}|< 0.9$, and $q=0.498_{-0.189}^{+0.055}$ 
for $|\chi_{i,z}|< 0.99$. While there is some overlap in 90\% confidence interval for 
the effective spin posteriors the peak parameters are not aligned likely due to 
many of the peak $\lnL$ simulations having high z-component spin, with our inferred 
posteriors of $\chieff=0.149_{-0.027}^{+0.206}$, $\chieff=0.387_{-0.116}^{+0.243}$,
and $\chieff=0.253_{-0.074}^{+0.363}$ respectively for $|\chi_{i,z}|<0.5$, $<0.9$, 
and $<0.99$. The chirp mass posteriors lie in the same region of parameter space with 
some marginal disagreement between peak values where 
$\mc=117.462_{-6.635}^{+13.882}\ M_\odot$,
$\mc=123.154_{-10.087}^{+10.984}\ M_\odot$, and 
$\mc=111.896_{-3.391}^{+16.187}\ M_\odot$ for $|\chi_{i,z}|<0.5$, $<0.9$, and $<0.99$ 
respectively. Finally the eccentricity posterior favors a moderate value with good 
agreement between the choice of spin prior, where $e_{20}=0.161_{-0.158}^{+0.045}$, 
$e_{20}=0.155_{-0.155}^{+0.015}$, and $e_{20}=0.190_{-0.189}^{+0.010}$ respectively 
for $|\chi_{i,z}|<0.5$, $<0.9$, and $<0.99$. The observations on our eccentricity 
posterior for the nonprecessing analysis can be seen in the top panel of 
\figref{fig:231123_e20_kde}. However the eccentric Bayes factors for each of the spin 
priors are all negative, indicating that this nonprecessing analysis prefers a 
quasicircular waveform where $\bayesecc=-0.82$, $\bayesecc=-0.93$, and $\bayesecc=-0.61$ 
respectively for $|\chi_{i,z}|<0.5$, $<0.9$, and $<0.99$. A summary of the maximum 
posterior values found through a KDE of all parameters in the nonprecessing analysis 
from the top panel of \figref{fig:231123_e20_PE} are given in 
\Tabref{tab:231123_e20kde_params} along with the quasicircular vs eccentric Bayes 
factors.

\begin{table}
\renewcommand{\arraystretch}{1.5}
  \caption{Summary of Maximum Kernel Density Estimates for the nonprecessing analysis 
  of GW231123 at $\fref = 20$ Hz as shown in the top panel of 
  \figref{fig:231123_e20_PE} and Bayes factors for the quasi-circular vs eccentric 
  hypothesis calculated from \Eqnref{eq:bayes}. The $e_{20}$ KDE is shown in the top 
  panel of \figref{fig:231123_e20_kde}.}
  \label{tab:231123_e20kde_params} 
\begin{ruledtabular}
\begin{tabular}{c|ccc}
Parameters & $|\chi_{i,z}|< 0.5$ & $|\chi_{i,z}|< 0.9$ & $|\chi_{i,z}|< 0.99$ \\  
  & posterior & posterior & posterior \\
\hline 
$\mc (M_\odot)$ & $117.462_{-6.635}^{+13.882}$ & $123.154_{-10.087}^{+10.984}$ & $111.896_{-3.391}^{+16.187}$  \\
$q$ & $0.499_{-0.010}^{+0.084}$ & $0.496_{-0.166}^{+0.102}$ & $0.498_{-0.189}^{+0.055}$ \\
$e_{20}$ & $0.161_{-0.158}^{+0.045}$ & $0.155_{-0.155}^{+0.015}$ & $0.190_{-0.189}^{+0.010}$\\
$\chieff$ & $0.149_{-0.027}^{+0.206}$ & $0.387_{-0.116}^{+0.243}$ & $0.253_{-0.074}^{+0.363}$ \\
$\bayesecc$ & -0.82 & -0.93 & -0.61 \\
\end{tabular}
\end{ruledtabular}
\end{table}

Next we focus on the precessing analysis shown in the bottom panel of 
\figref{fig:231123_e20_PE}, which adds precessing simulations to the grid used 
in the nonprecessing analysis. Looking at the marginal likelihoods generated from 
direct comparison of each NR simulation to the data, the best matching simulations 
are noneccentric at 20 Hz with all except one in the top ten by $\lnL$ having 
negligible eccentricity, many have moderate positive effective spin, large in-plane spin of the 
primary mass $\chi_{1,\perp}$, and mass ratios split evenly across $q\approx 0.7$ and 
equal mass, as seen by the top ten matching simulations by $\lnL$ in 
\Tabref{tab:231123_e20_prec_peakLnL}. The best fitting simulation of the precessing 
analysis, RIT:BBH:1986, has mildly asymmetric masses with $q=0.730$, effective spin 
$\chieff = 0.210$, in-plane spin of the primary mass $\chi_{1,\perp}=0.718$, and zero 
eccentricity at 20 Hz with a $\lnL=220.701$ more than 10 higher than the 
nonprecessing best of $\lnL=210.591$ for RIT:BBH:1816. All ten of the best matching 
simulations from our precessing analysis in \tabref{tab:231123_e20_prec_peakLnL} 
have likelihoods higher than those of the nonprecessing analysis presented in 
\Tabref{tab:231123_e20_aligned_peakLnL}. We note here that the top $\lnL$ NR 
simulation RIT:BBH:1986 in both Tables \ref{tab:231123_e20_prec_peakLnL} and 
\ref{tab:231123_e10_prec_peakLnL} corresponds to a primary black hole with intrinsic 
spins $s_1=0.95$ and secondary $s_2=0.36$ in a highly precessing noneccentric 
configuration.

\begin{table*} 
  \caption{Peak marginal likelihood, $\lnLmarg$, NR simulations 
  of GW231123 for the $\fref=20$ Hz precessing analysis shown in the bottom panel of
  \figref{fig:231123_e20_PE}. The likelihood values reported in the  
  $\lnL$ column are a result of Gaussian process fitting discussed in 
  \secref{s:events}. $\frefnr$ is the initial apoapsis NR frequency, which is used in the conversion of eccentricity to an apoapsis frequency of $\fref=20$ Hz. 
  $e_{20}$ is the conversion of the initial 3.5PN order eccentricity, 
  $e_{3.5\rm{PN}}$, to 20 Hz using \eqnref{eq:e20}.}
  \label{tab:231123_e20_prec_peakLnL} 
\begin{ruledtabular}
\begin{tabular}{cccccccccccccccc}
NR\_sim\_id	& $m_1/M_\odot$& $m_2/M_\odot$&	$q$& $s_1^x$& $s_1^y$& $s_1^z$& $s_2^x$& $s_2^y$& $s_2^z$& $e_{3.5\rm{PN}}$& $e_{20}$& $\lnL$& $\frefnr$\\
 & & & & & & & & & & & & & (Hz)\\
\hline
RIT:BBH:1986& 178.492& 130.262& 0.730& -0.709& 0.109& 0.624& 0.022& 0.065& -0.357& 0.000& 0.000& 220.701& 4.43\\
RIT:BBH:0272& 188.337& 125.611& 0.667& -0.150& 0.675& 0.405& 0.000& 0.000& 0.000& 0.002& 0.000& 218.657& 4.93\\
RIT:BBH:0977& 149.875& 144.814& 0.966& -0.539& 0.245& -0.004& -0.216& 0.709& 0.144& 0.005& 0.002& 218.431& 7.65\\
RIT:BBH:0969& 148.786& 148.784& 1.000& -0.554& 0.098& 0.570& -0.565& 0.119& -0.555& 0.002& 0.000& 218.093& 5.02\\
RIT:BBH:0271& 186.050& 124.089& 0.667& 0.207& 0.664& 0.397& 0.000& 0.000& 0.000& 0.002& 0.000& 217.719& 5.00\\
RIT:BBH:1633& 147.319& 147.319& 1.000& 0.700& 0.000& 0.000& 0.700& 0.000& 0.000& 0.445& 0.076& 217.705& 2.08\\
RIT:BBH:0627& 186.534& 124.409& 0.667& 0.316& -0.609& 0.413& 0.000& 0.000& 0.000& 0.002& 0.000& 217.432& 5.06\\
RIT:BBH:0130& 173.848& 130.394& 0.750& 0.492& 0.191& 0.602& 0.000& 0.000& -0.801& 0.001& 0.000& 217.409& 6.71\\ 
RIT:BBH:0989& 160.573& 160.554& 1.000& -0.241& 0.763& 0.008& 0.004& -0.012& 0.801& 0.001& 0.000& 217.380& 4.60\\
RIT:BBH:1010& 161.215& 161.210& 1.000& -0.782& 0.172& 0.009& 0.006& -0.010& 0.801& 0.002& 0.000& 217.317& 4.58\\

\end{tabular}
\end{ruledtabular}
\end{table*}

\begin{figure*}[h!tbp]
    \includegraphics[width=\textwidth]{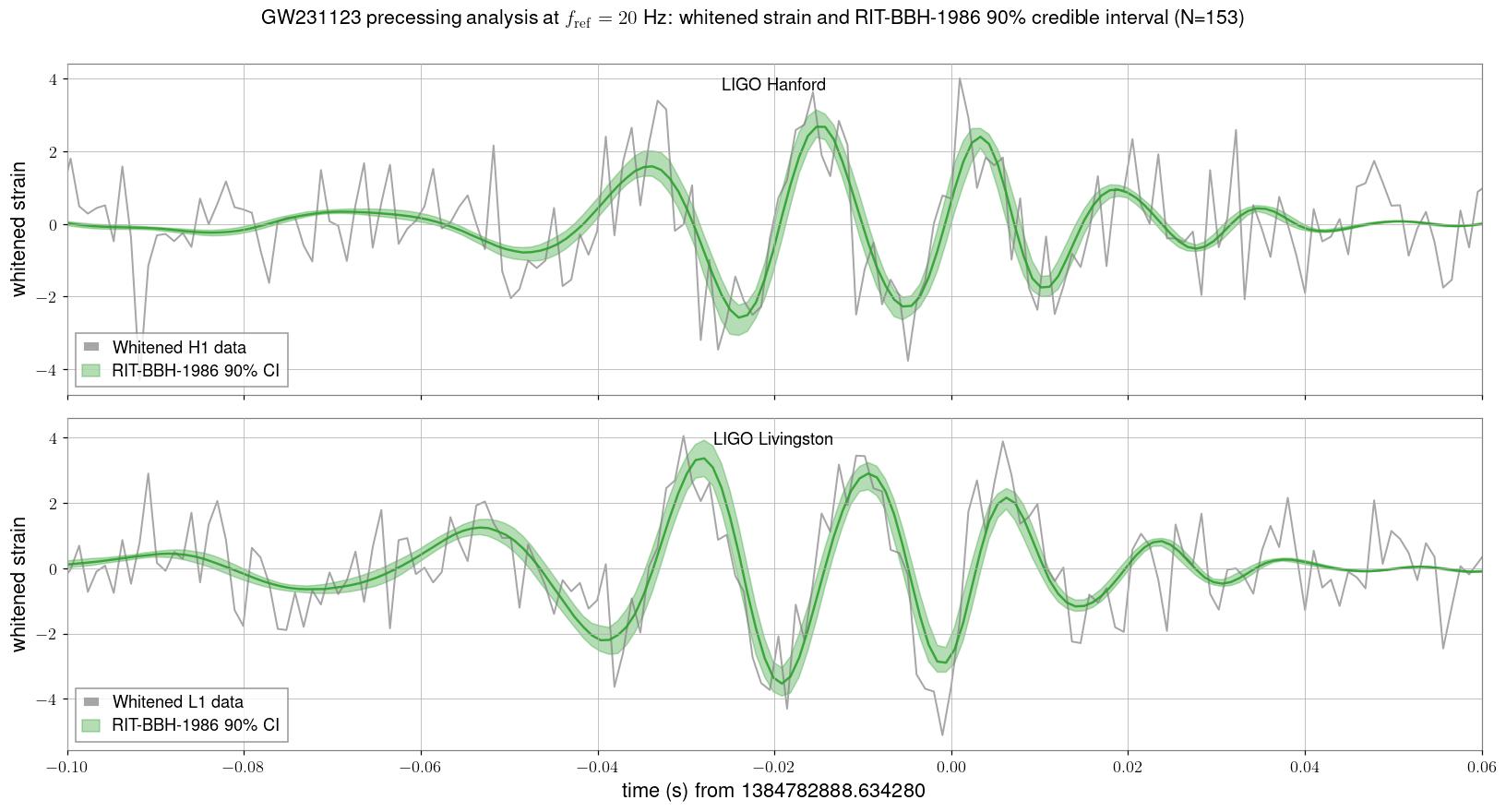}
    \caption{Best fitting time-domain NR waveform RIT:BBH:1986 median (dark green) and 90\% credible interval (light green band) overlaid on whitened strain data (grey) near GW231123.}
    \label{fig:231123_e20_prec_wf}
\end{figure*}

\begin{figure}[h!tbp]
    \includegraphics[width=0.46\textwidth]{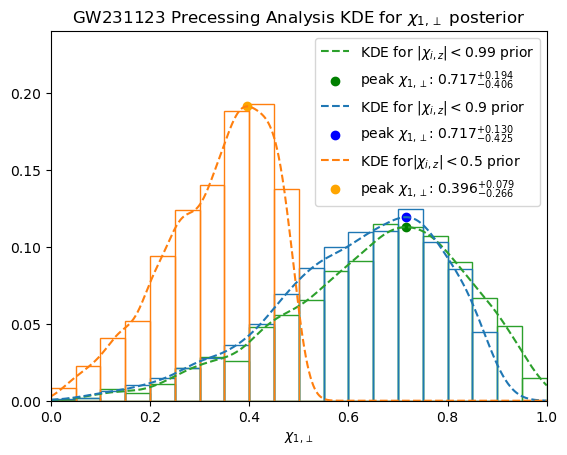}
    \caption{Kernel density estimate of the GW231123 NR-based PE 
    1-dimensional $\chi_{1,\perp}$ posterior at 20\,Hz from the aligned plus 
    precessing spin analysis (histograms from bottom panel of \figref{fig:231123_e20_PE}) for 
    the three different spin priors, $|\chi_{i,z}|< 0.99$ (green), $|\chi_{i,z}|< 0.9$ 
    (blue) and $|\chi_{i,z}|< 0.5$ (orange). The maximum of the KDE for each prior is 
    indicated by a dot of the same color with errors reported as the 90\% confidence 
    interval of the histogram.}
    \label{fig:231123_chi1perp_kde}
\end{figure}

These sparsely sampled marginal likelihoods are interpolated over mass, aligned spins, 
eccentricity, and the in-plane transverse spins $\chi_{i,\perp}$ generating an 
inferred joint posterior over these parameters. The mass ratio found in the precessing 
analysis falls higher than the nonprecessing case, with values closer to equal mass 
$q=0.948_{-0.332}^{+0.040}$, $q=0.946_{-0.297}^{+0.042}$, and 
$q=0.945_{-0.297}^{+0.041}$ respectively for $|\chi_{i,z}|<0.5$, $<0.9$, 
and $<0.99$. We find similar disagreement between the precessing versus 
nonprecessing case for both effective spin and chirp mass. The effective spins are 
lower with values of $\chieff=0.047_{-0.254}^{+0.188}$, 
$\chieff=0.063_{-0.300}^{+0.320}$, and $\chieff=0.121_{-0.368}^{+0.289}$ respectively 
for $|\chi_{i,z}|<0.5$, $<0.9$, and $<0.99$. Chirp masses are higher
with wider 90\% confidence intervals over the choice of spin prior with 
$\mc=150.856_{-39.131}^{+38.893}\ M_\odot$, 
$\mc=157.813_{-43.246}^{+32.094}\ M_\odot$, 
and $\mc=175.649_{-59.666}^{+14.652}\ M_\odot$ respectively for $|\chi_{i,z}|<0.5$, 
$<0.9$, and $<0.99$. The $\chi_{1,\perp}$ posterior (shown in 
\figref{fig:231123_chi1perp_kde}) favors large transverse spin with
$\chi_{1,\perp}=0.396_{-0.266}^{+0.079}$, $\chi_{1,\perp}=0.717_{-0.425}^{+0.130}$,
and $\chi_{1,\perp}=0.717_{-0.406}^{+0.194}$ for $|\chi_{i,z}|<0.5$, $<0.9$, 
and $<0.99$ respectively. We can clearly see in 
\figref{fig:231123_chi1perp_kde} that the posterior generated with the 
$|\chi_{i,z}|<0.5$ prior is dominated by the prior with railing evident at boundary.
The precessing Bayes factors found for each choice of spin 
prior are very large and positive confirming our observations on the transverse spin 
posteriors and suggesting this event prefers spin precession, where 
$\bayesprec=12.17$, $\bayesprec=11.24$, and $\bayesprec=10.93$ for 
$|\chi_{i,z}|<0.5$, $<0.9$, and $<0.99$, respectively.
Finally, the eccentricity posterior clearly peaks near zero 
with all three spin priors agreeing on a peak near $e_{20}\sim 0.06$ where 
$e_{20}=0.061_{-0.047}^{+0.071}$ for $|\chi_{i,z}|<0.5$, 
$e_{20}=0.058_{-0.047}^{+0.061}$ for $|\chi_{i,z}|<0.9$, and 
$e_{20}=0.056_{-0.045}^{+0.059}$ for $|\chi_{i,z}|<0.99$. 
We see from the eccentricity KDE in the bottom panel of \figref{fig:231123_e20_kde} 
that there is significant support for eccentricity near zero.
The eccentric Bayes factors clearly support the conclusion that the precessing 
analysis favors a quasi-circular waveform as all three Bayes factors are  
negative with $\bayesecc=-0.48$ for $|\chi_{i,z}|<0.5$, and $\bayesecc=-0.64$ for 
both $|\chi_{i,z}|<0.9$ and $<0.99$. The posterior values and Bayes 
factors are summarized in \Tabref{tab:231123_e20prec_kde_params} for the precessing 
analysis of GW231123.

\begin{table}
\renewcommand{\arraystretch}{1.5}
  \caption{Summary of maximum kernel density estimates for the precessing analysis of GW231123 at $\fref = 20$ Hz as shown in the bottom panel of \figref{fig:231123_e20_PE}. Bayes factors for the quasi-circular vs eccentric hypothesis, $\bayesecc$, calculated from \Eqnref{eq:bayes} and the nonprecessing vs precessing hypothesis, $\bayesprec$, calculated from \eqnref{eq:bayesprec}. The $e_{20}$ KDE is shown in the bottom panel of \figref{fig:231123_e20_kde}. The $\chi_{1,\perp}$ KDE is shown in \figref{fig:231123_chi1perp_kde}.} 
  \label{tab:231123_e20prec_kde_params} 
\begin{ruledtabular}
\begin{tabular}{c|ccc}
Parameters & $|\chi_{i,z}|< 0.5$ & $|\chi_{i,z}|< 0.9$ & $|\chi_{i,z}|< 0.99$ \\  
  & posterior & posterior & posterior \\
\hline 
$\mc (M_\odot)$ & $150.856_{-39.131}^{+38.893}$ & $157.813_{-43.246}^{+32.094}$ & $175.649_{-59.666}^{+14.652}$  \\
$q$ & $0.948_{-0.332}^{+0.040}$ & $0.946_{-0.297}^{+0.042}$ & $0.945_{-0.297}^{+0.041}$ \\
$e_{20}$ & $0.061_{-0.047}^{+0.071}$ & $0.058_{-0.047}^{+0.061}$ & $0.056_{-0.045}^{+0.059}$\\
$\chieff$ & $0.047_{-0.254}^{+0.188}$ & $0.063_{-0.300}^{+0.320}$ & $0.121_{-0.368}^{+0.289}$ \\
$\chi_{1, \perp}$ & $0.396_{-0.266}^{+0.079}$ & $0.717_{-0.425}^{+0.130}$ & $0.717_{-0.406}^{+0.194}$ \\ 
$\bayesecc$ & -0.48 & -0.64 & -0.64\\
$\bayesprec$ & 12.17 & 11.24 & 10.93 \\
\end{tabular}
\end{ruledtabular}
\end{table}

In our analysis of GW231123 in both a nonprecessing and a precessing regime, we again find
higher likelihoods in the precessing analysis as well as more tightly constrained 
parameters especially in the case of mass ratio and effective spin $\chieff$. This 
event is clearly better matched to a precessing noneccentric waveform where the 
inclusion of spins up to their maximal value is vital to correctly estimating mass 
ratio $q$, $\chieff$, and $\chi_{1,\perp}$. As supported by the  
negative Bayes factor of $\bayesecc=-0.64$ for the posterior generated with the 
$|\chi_{i,z}|<0.99$ prior in the precessing analysis which indicates support for a 
quasi-circular waveform as well as a positive precessing Bayes factor suggesting this 
event prefers spin-precession. 
We choose the posterior generated with the $|\chi_{i,z}|<0.99$ prior as our preferred 
posterior which is motivated by the need to include the maximal 
number of simulations when generating our posterior. Our finding 
that the GW231123 $e_{20}$ posterior peaks at nonzero eccentricity 
(albeit with negative Bayes factors) when spin precession is neglected is consistent 
with injection-recovery studies where a non-eccentric precessing signal injection is 
recovered with a non-precessing eccentric waveform as done in 
\cite{RomeroShaw_2020,Romero_Shaw_2023,Divyajyoti_2024,Divyajyoti_2026} but when we 
include spin precession the eccentricity vanishes.

We can compare our preferred results from the precessing analysis using the 
$|\chi_{i,z}|<0.99$ spin prior to the noneccentric results from the LVK
\cite{GW231123_special}. We find a larger mass ratio and larger chirp mass in the 
detector frame, smaller effective spin, and similar in-plane spin for the primary 
mass\footnote{The chirp mass in the detector frame and in-plane spin for the primary 
mass $\chi_{1,\perp}$ were retrieved from the associated data release 
\cite{GW231123_specialdatarelease} for the LVK analysis \cite{GW231123_special}.}
where our (LVK) median values are $q=0.945$ ($q=0.74$), $\mc=175.649 \ M_\odot$ 
($\mc=137\ M_\odot$), $\chieff=0.121$ ($\chieff=0.32$), and $\chi_{1,\perp}=0.717$ 
($\chi_{1,\perp}=0.74$) noting that the LVK analysis dealt with significant waveform 
systematics that result in having no reliable determination of the spin orientation. 
Additionally the LVK analysis \cite{GW231123_special} find $\chi_{\rm p}=0.77$ as this 
value is the standard effective spin precession parameter used in the literature 
values compared below. From Malagon \& O'Shaughnessy \cite{malagon_2026_RIFTecc} the 
median values from Table 1 in \cite{malagon_2026_RIFTecc} for detector frame chirp 
mass and mass ratio are lower than our values, where they find $\mc=145.72 \ M_\odot$ 
and $q=0.689$ (calculated from the median values of $m_1$ and $m_2$), effective spin 
and eccentricity are much different with $\chieff=0.58$ and $e_{13.33\rm Hz}=0.46$ 
respectively. We note that Ref~\cite{malagon_2026_RIFTecc} do not use waveforms that 
include spin precession simultaneously with eccentricity but Jan {\it et al.} 
\cite{jan2026_gw231123} does perform such an analysis. The reported median parameters 
for the eccentric and precessing analysis from Jan {\it et al.} 
\cite{jan2026_gw231123} find $q=0.69$, $\mc= 155.6\ M_\odot$, $\chieff=0.43$, 
$\chi_{\rm p}=0.71$, and $e_{10}=0.062$. Our value of eccentricity is most similar 
to Ref~\cite{jan2026_gw231123}, while the remaining parameters show some disagreement.
Finally we compare to the quasicircular precessing results found with the 
IMRPhenomTPHM waveform from Xu {\it et al.} \cite{xu2025_o4a_ecc}, although they do 
not provide exact median parameters the key values can be estimated from Figure 3 in 
\cite{xu2025_o4a_ecc}, they find $q\approx0.8$, $\chieff\approx0.25$, 
$\chi_{\rm p}\approx0.7$, and $\mc\approx150 \ M_\odot$ (where 
$M\approx 350 \ M_\odot$). Based on the loose approximations from 
\cite{xu2025_o4a_ecc} our values for mass ratio and chirp mass are higher, while our 
effective spin is lower, and the effective spin precession $\chi_{\rm p}$ is not 
directly comparable to our $\chi_{1,\perp}$. 

We present in \figref{fig:231123_e20_prec_wf} a waveform reconstruction of our best 
matching simulation by likelihood, RIT:BBH:1986, along with the band-limited whitened 
strain data near the GW231123 detector response in the LIGO Hanford (H1) and LIGO 
Livingston (L1) instruments. We achieved 153 fair-draw samples for our peak likelihood 
noneccentric precessing simulation, which has $\lnL=220.701$, $q=0.730$, $\chieff=0.21$, 
$\chi_{1,\perp}=0.718$, and $e_{20}=0$. Our reconstruction fits the LIGO strain 
data very well in both amplitude and phase, corroborating the likelihood value we 
achieve and shows that this simulation is truly a good match to the data.

\subsection{Recoil estimates for GW190521 and GW231123}
\label{sec:recoil-events}
We note here that the two clear top matching precessing simulations for GW190521, RIT:BBH:1618 and RIT:BBH:1616 in \Tabref{tab:190521_prec_peakLnL} have a computed 
\cite{Ficarra:2026cej} magnitude of the recoil velocity of 3965\,km/s.

We also note that the clear top matching precessing simulation, RIT:BBH:1986 for GW231123 in \Tabref{tab:231123_e20_prec_peakLnL} has a computed 
\cite{Ficarra:2026cej} magnitude of the recoil velocity of 2059\,km/s.

While those values are not computed through, for instance, a KDE estimation process, 
and are properties of the best-fit simulations, we highlight here
that those simulations clearly stand out above the next in $\lnL$ values by a $\Delta\lnL\approx1.7$ for GW190521 and $\Delta\lnL\approx2$ for GW231123 corresponding to a $\approx2\sigma$ (a 95\% confidence interval for the single recoil magnitude parameter).

\section{Conclusions and discussion}
\label{s:conclusion}

From our direct comparison of NR simulations to GW events we conclude that we can 
simultaneously measure eccentricity and spin-induced precession in gravitational wave 
events and we are able to distinguish between those two effects with our techniques.
In fact, for the case of GW190521 there is a clear preference for spin precession when 
looking at the top likelihood values in \Tabref{tab:190521_prec_peakLnL} 
where the peak is near $\lnL\approx95$ compared to the
nonprecessing cases in \Tabref{tab:190521_aligned_peakLnL} with a peak near 
$\lnL\approx92$. Notably, both the precessing and nonprecessing top simulations 
prefer an eccentricity at 20 Hz of $e_{20}\sim0.15$.

Similarly, for GW231123, there is a clear preference for spin precession when looking 
at the top likelihood values in \Tabref{tab:231123_e20_prec_peakLnL} with a peak of 
$\lnL\approx220$ compared to the nonprecessing cases in 
\Tabref{tab:231123_e20_aligned_peakLnL} which peak near $\lnL\approx210$. This
contrasts with GW190521, as when we consider eccentricity for the nonprecessing case, 
preferred values lie near $e_{20}\sim0.15$, but for the precessing simulations it 
clearly favors $e_{20}\sim0$.

A more detailed statistical analysis quantitatively confirms the above account (which was based on single simulations) and is provided by our parameter estimates found in 
the kernel density estimate Tables \ref{tab:190521_kde_params} and 
\ref{tab:190521_prec_kde_params}, as well as \figref{fig:190521_e20_kde}, for 
GW190521; and Tables \ref{tab:231123_e20kde_params} and 
\ref{tab:231123_e20prec_kde_params}, as well as the corresponding Figure 
\ref{fig:231123_e20_kde}, for GW231123. This reinforces our conclusion that we can 
differentiate the effects of spin precession and eccentricity, with two clear examples 
of the application of our technique to GW190521 which prefers precession and 
eccentricity while GW231123 prefers precession but not eccentricity upon entering 
the LVK sensitivity band. These claims are further corroborated by the 
eccentric and precessing Bayes factors, where we find a positive Bayes factor for both 
eccentricity and precession for GW190521 but for GW231123 we find a negative eccentric 
Bayes factor and a positive precessing Bayes factor.

In this regard, we also point out that we followed the pattern of our previous paper 
\cite{McMillin:2025hof} where we considered priors at successive values of the aligned 
spin $|\chi_{i,z}|<0.5$ and $|\chi_{i,z}|<0.9$, where our previous
paper used these priors to make comparisons with phenomenological waveform models. 
Here we have used an additional prior that included all our simulations up to 
$|\chi_{i,z}|<0.99$ and they proved to play a crucial role since looking at our peak 
likelihood NR simulations in both Tables \ref{tab:190521_prec_peakLnL} and 
\ref{tab:231123_e20_prec_peakLnL}, for the two events GW190521 and GW231123 analyzed 
in this paper, the consideration of large spins with $s>0.90$ is crucial,
and they contribute to a better estimate of the binaries parameters as displayed in 
the KDE Tables \ref{tab:190521_prec_kde_params}, \ref{tab:231123_e20prec_kde_params}, 
and Figures \ref{fig:190521_e20_kde}, \ref{fig:231123_e20_kde}, 
and \ref{fig:231123_chi1perp_kde}.

We also mention here that when analyzing GW231123, we also considered a low frequency 
cutoff of 10 Hz, as we display in the \appendref{s:231123_e10}. Our findings for this 
event are that this extension of the sensitivity band does not produce qualitative 
changes in any of our results evaluated at 20 Hz and thus we conclude that they are 
robust against the inclusion of signal contributions down to 10 Hz.

Finally we mention that the large recoil values obtained for GW190521 and GW231123 
from the top likelihood simulations are due to the presence of both high spin 
components and high precession that lead to relatively large recoil probabilities for 
comparable mass binaries \cite{Lousto:2009ka,Lousto:2012gt}.
Hence the outcome of the recoil magnitudes of the order of:
$|V_{recoil}|\simeq4\times10^{3}$~km\,s$^{-1}$ for GW190521 and
$|V_{recoil}|\simeq2\times10^{3}$~km\,s$^{-1}$ for GW231123.

Considering the results of this paper along with our previous work in 
\cite{McMillin:2025hof} we are confident that parameter estimation of gravitational 
wave events using our ever growing bank of numerical relativity simulations will 
be key to analyzing additional LVK events, especially those that may harbor
eccentricity and/or precession.


\begin{acknowledgements}
This material is based upon work supported by NSF's LIGO Laboratory which is a 
major facility fully funded by the National Science Foundation.
GF gratefully acknowledges the support of University of Calabria through a research fellowship funded by DR 1688/2023.
COL gratefully acknowledges support from NSF awards AST-2319326, PHY-2207920 and PHY-2513442.
ROS gratefully acknowledges support from NSF awards NSF PHY-1912632, 
PHY-2012057, PHY-2309172, AST-2206321, and the Simons Foundation. Finally, the authors thank James Healy for support on using the RIT Catalog of BBH simulations.
\end{acknowledgements}

\appendix

\section{GW231123 at $\fref = 10$ Hz}
\label{s:231123_e10}
Here we provide an equivalent analysis of GW231123 as presented in 
\secref{ss:gw231123} but with a low frequency cutoff $\flow$ of 10 Hz and $\fref$ 
of 10 Hz. We find that this change in frequency band has minimal effects on the 
results found with a 20 Hz cutoff. In particular we find nearly the same peak 
likelihoods for our top simulations where the 10 Hz (20 Hz) analysis returns 
$\lnL = 210.581$ ($\lnL = 210.591$) for the nonprecessing RIT:BBH:1816 and 
$\lnL=220.686$ ($\lnL=220.701$) for the precessing RIT:BBH:1986, a similar minor 
decrease in $\lnL$ from the 20 Hz analysis to the 10 Hz analysis is seen in all 
simulations in Tables \ref{tab:231123_e10_aligned_peakLnL}, and \ref{tab:231123_e10_prec_peakLnL}.

\begin{figure*}[h!tbp]
    \includegraphics[width=1.45\columnwidth]{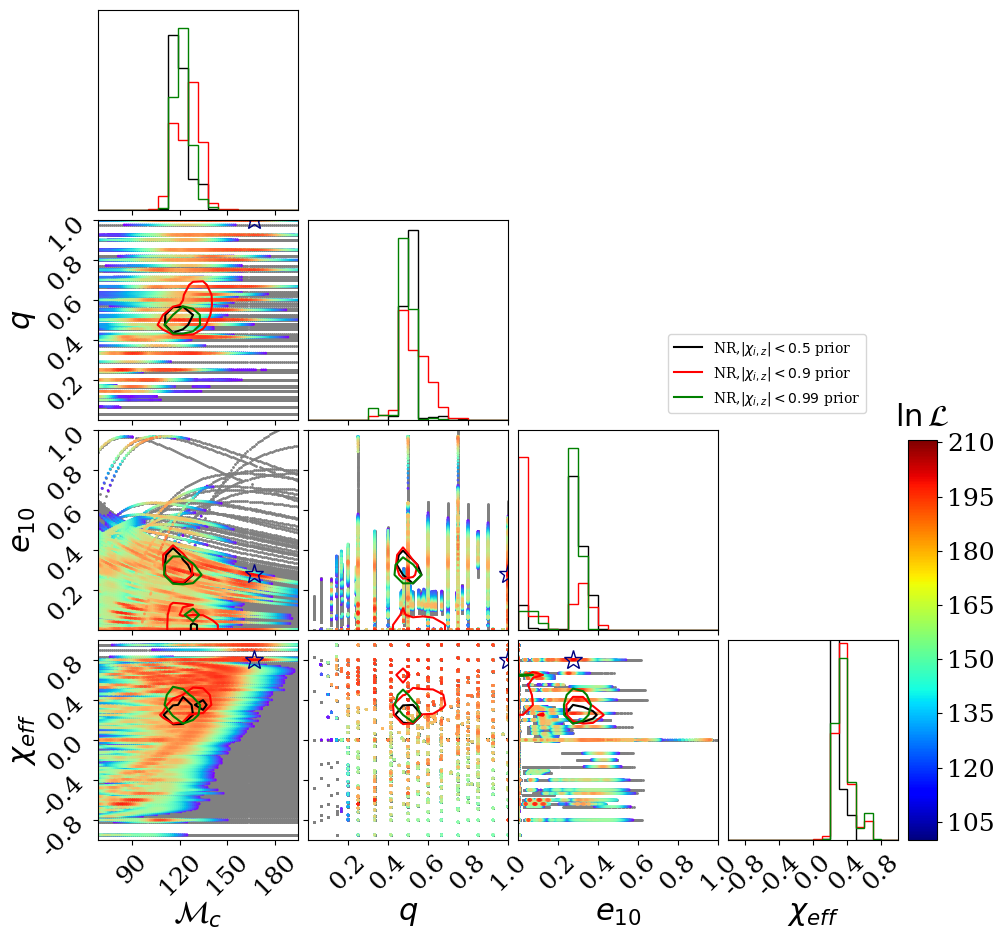}
    \includegraphics[width=1.45\columnwidth]{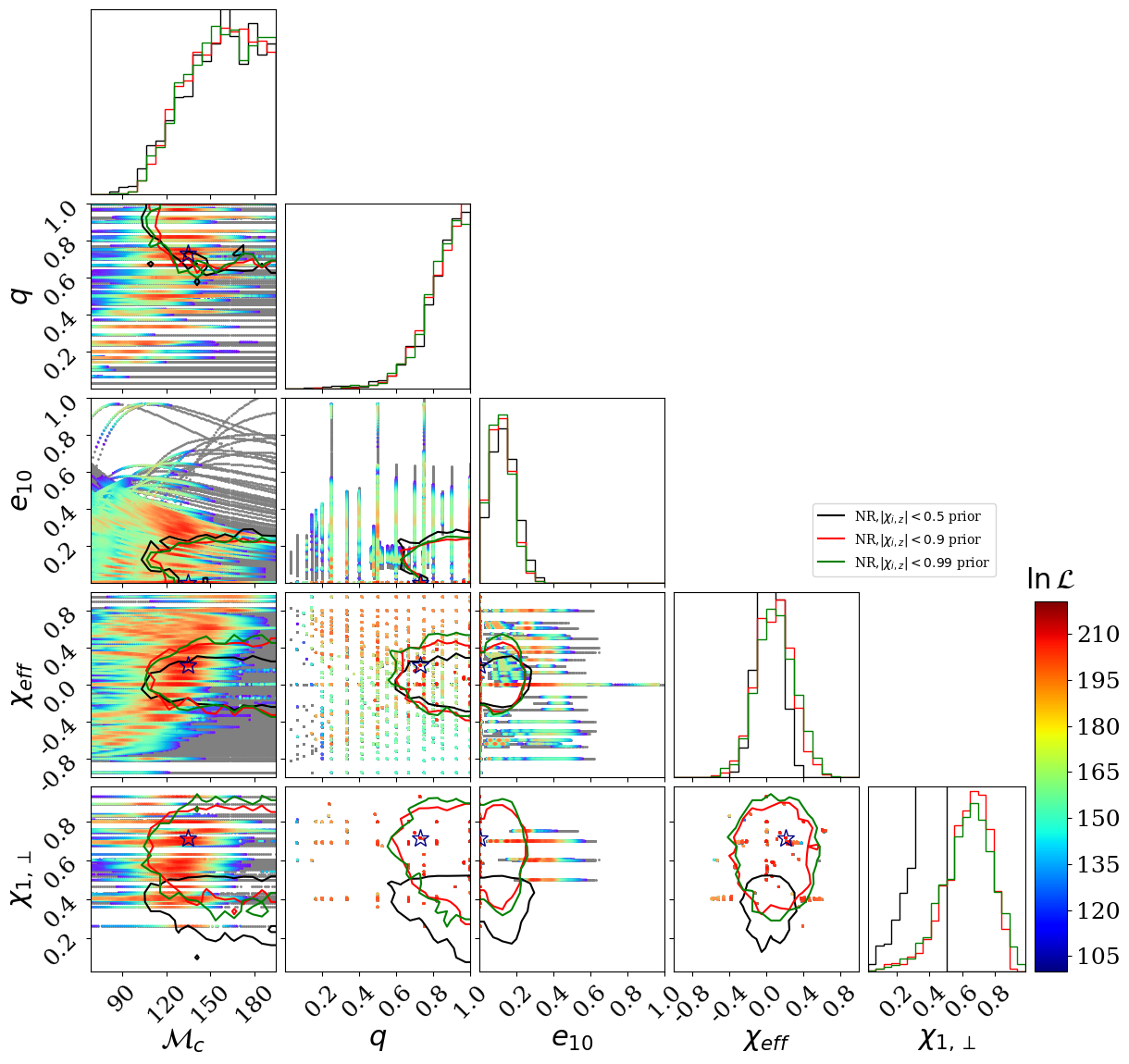}
    \caption{A corner plot showing the results of NR based PE for 
    GW231123 with a grid consisting of the RIT catalog \cite{Ficarra:2026cej} 
    with the restrictions on mass ratio from \secref{ss:analysis} for a total 
    of 1519 simulations for the nonprecessing analysis with  
    $\fref=10$ Hz ({\it top}) and 2117 simulations for the precessing analysis with  
    $\fref=10$ Hz ({\it bottom}). These are rescaled over the detector frame chirp 
    mass range of $\mc \in [68.2,194.6] \ M_\odot$. Three NR posteriors are shown with 
    different $|\chi_{i,z}|$ priors, $|\chi_{i,z}|< 0.99$ (green), $|\chi_{i,z}|< 0.9$ 
    (red), and $|\chi_{i,z}|< 0.5$ (black).}
    \label{fig:231123_e10_PE}
\end{figure*}

Looking at the nonprecessing analysis in the top panel and the precessing analysis 
in the bottom panel of \figref{fig:231123_e10_PE} we can draw almost identical 
conclusions as discussed in \secref{ss:gw231123} noting that the $e_{10}$ posterior 
is not directly comparable to the $e_{20}$ posterior since they have differing 
reference frequencies. However, the same general trend of a nonzero eccentricity 
for the nonprecessing case is seen matching the 20 Hz nonprecessing analysis.
We maintain negative eccentric Bayes factors indicating that both the 
nonprecessing and precessing analyses prefer a quasicircular waveform. 
When we look at the KDE Tables \ref{tab:231123_e10kde_params} and 
\ref{tab:231123_e10prec_kde_params} there is some spread in the peak parameters from 
the 20 Hz analysis however the 90\% confidence intervals have significant overlap. We 
continue to see a convergence in the mass ratio by adding the precessing 
simulations. Overall the differences we see between the 20 Hz and 10 Hz analysis are 
very minor, and an observed minor decrease in $\lnL$ for the peak simulations in the 
10 Hz analysis suggest that the 20 Hz analysis is preferred.

\begin{table*}[hbtp] 
  \caption{Peak marginal likelihood, $\lnLmarg$, NR simulations 
  of GW231123 for the $\fref=10$ Hz nonprecessing analysis shown in the top panel of
  \figref{fig:231123_e10_PE}. Where all have the dimensionless spin components 
  $s_1^x=s_1^y=s_2^x=s_2^y=0$. The likelihood values reported in the  
  $\lnL$ column are a result of Gaussian process fitting discussed in 
  \secref{s:events}. $\frefnr$ is the initial apoapsis NR frequency, which is used in the conversion of eccentricity to an apoapsis frequency of $\fref=10$ Hz. 
  $e_{10}$ is the conversion of the initial 3.5PN order eccentricity, 
  $e_{3.5\rm{PN}}$, to 10 Hz using \eqnref{eq:e20}.}
  \label{tab:231123_e10_aligned_peakLnL} 
\begin{ruledtabular}
\begin{tabular}{cccccccccccccc}

NR\_sim\_id	&	$m_1/M_\odot$&	$m_2/M_\odot$&	$q$&	$s_1^z$&	$s_2^z$&	$e_{3.5\rm{PN}}$&  $e_{10}$&  $\lnL$& $\frefnr$ (Hz)\\
\hline
RIT:BBH:1816& 191.306& 191.306& 1.000& 0.801& 0.801& 0.803& 0.277& 210.581& 1.56\\
RIT:BBH:1859& 200.698& 50.204& 0.250& -0.801& 0.000& 0.915& 0.085& 205.866& 0.39\\
RIT:BBH:1724& 212.904& 106.454& 0.500& 0.801& -0.801& 0.714& 0.278& 205.326& 1.90\\
RIT:BBH:0796& 235.865& 118.046& 0.500& 0.952& 0.000& 0.001& 0.001& 204.854& 5.33\\
RIT:BBH:1721& 196.679& 98.340& 0.500& 0.400& 0.400& 0.717& 0.296& 204.716& 2.02\\
RIT:BBH:1880& 184.359& 61.489& 0.334& -0.801& 0.000& 0.916& 0.108& 204.436& 0.49\\
RIT:BBH:1915& 262.547& 87.554& 0.333& 0.950& -0.950& 0.000& 0.000& 204.308& 5.36\\
RIT:BBH:1817& 186.907& 186.907& 1.000& 0.801& 0.801& 0.835& 0.275& 203.924& 1.45\\
RIT:BBH:0314& 233.794& 116.985& 0.500& 0.851& 0.500& 0.001& 0.000& 203.785& 3.64\\
RIT:BBH:1008& 228.323& 125.580& 0.550& 0.800& 0.800& 0.001& 0.001& 203.572& 4.46\\
\end{tabular}
\end{ruledtabular}
\end{table*}

\begin{table*} 
  \caption{Peak marginal likelihood, $\lnLmarg$, NR simulations 
  of GW231123 for the $\fref=10$ Hz precessing analysis shown in the bottom panel of
  \figref{fig:231123_e10_PE}. The likelihood values reported in the  
  $\lnL$ column are a result of Gaussian process fitting discussed in 
  \secref{s:events}. $\frefnr$ is the initial apoapsis NR frequency, which is used in the conversion of eccentricity to an apoapsis frequency of $\fref=10$ Hz. 
  $e_{10}$ is the conversion of the initial 3.5PN order eccentricity, 
  $e_{3.5\rm{PN}}$, to 10 Hz using \eqnref{eq:e20}.}
  \label{tab:231123_e10_prec_peakLnL} 
\begin{ruledtabular}
\begin{tabular}{cccccccccccccccc}
NR\_sim\_id	& $m_1/M_\odot$& $m_2/M_\odot$&	$q$& $s_1^x$& $s_1^y$& $s_1^z$& $s_2^x$& $s_2^y$& $s_2^z$& $e_{3.5\rm{PN}}$& $e_{10}$& $\lnL$& $\frefnr$\\
 & & & & & & & & & & & & & (Hz)\\
\hline
RIT:BBH:1986& 181.715& 132.614& 0.730& -0.709& 0.109& 0.624& 0.022& 0.065& -0.357& 0.000& 0.000& 220.686& 4.35\\
RIT:BBH:0272& 187.574& 125.103& 0.667& -0.150& 0.675& 0.405& 0.000& 0.000& 0.000& 0.002& 0.001& 218.640& 4.95\\
RIT:BBH:0977& 151.367& 146.256& 0.966& -0.539& 0.245& -0.004& -0.216& 0.709& 0.144& 0.005& 0.003& 218.410& 7.57\\
RIT:BBH:0969& 149.123& 149.121& 1.000& -0.554& 0.098& 0.570& -0.565& 0.119& -0.555& 0.002& 0.001& 218.077& 5.01\\
RIT:BBH:0271& 184.729& 123.208& 0.667& 0.207& 0.664& 0.397& 0.000& 0.000& 0.000& 0.002& 0.001& 217.702& 5.03\\
RIT:BBH:1633& 147.319& 147.319& 1.000& 0.700& 0.000& 0.000& 0.700& 0.000& 0.000& 0.445& 0.149& 217.680& 2.08\\
RIT:BBH:1021& 152.139& 151.297& 0.994& -0.517& 0.010& 0.660& -0.696& 0.286& -0.540& 0.002& 0.002& 217.423& 7.96\\
RIT:BBH:0989& 157.592& 157.573& 1.000& -0.241& 0.763& 0.008& 0.004& -0.012& 0.801& 0.001& 0.000& 217.422& 4.69\\
RIT:BBH:0627& 188.337& 125.611& 0.667& 0.316& -0.609& 0.413& 0.000& 0.000& 0.000& 0.002& 0.001& 217.412& 5.01\\
RIT:BBH:0959& 142.409& 142.408& 1.000& -0.416& 0.374& 0.573& -0.427& 0.379& -0.561& 0.002& 0.001& 217.365& 5.24\\

\end{tabular}
\end{ruledtabular}
\end{table*}

\begin{table}
\renewcommand{\arraystretch}{1.5}
  \caption{Summary of Maximum Kernel Density Estimates for the nonprecessing analysis
  of GW231123 at $\fref = 10$ Hz as shown in the top panel of \figref{fig:231123_e10_PE} and Bayes factors for the quasi-circular vs eccentric 
  hypothesis calculated from \Eqnref{eq:bayes}. the $e_{10}$ KDE is shown in the top panel of \figref{fig:231123_e10_kde}.} 
  \label{tab:231123_e10kde_params} 
\begin{ruledtabular}
\begin{tabular}{c|ccc}
Parameters & $|\chi_{i,z}|< 0.5$ & $|\chi_{i,z}|< 0.9$ & $|\chi_{i,z}|< 0.99$ \\  
  & posterior & posterior & posterior \\
\hline 
$\mc (M_\odot)$ & $117.588_{-2.544}^{+15.482}$ & $128.720_{-15.911}^{+7.168}$ & $118.980_{-3.320}^{+11.077}$  \\
$q$ & $0.502_{-0.017}^{+0.032}$ & $0.497_{-0.071}^{+0.154}$ & $0.497_{-0.130}^{+0.010}$ \\
$e_{10}$ & $0.292_{-0.290}^{+0.076}$ & $0.004_{-0.003}^{+0.381}$ & $0.294_{-0.290}^{+0.019}$\\
$\chieff$ & $0.271_{-0.025}^{+0.132}$ & $0.369_{-0.138}^{+0.221}$ & $0.387_{-0.137}^{+0.227}$ \\
$\bayesecc$ & -0.98 & -1.13 & -1.05 \\
\end{tabular}
\end{ruledtabular}
\end{table}

\begin{table}[h!]
\renewcommand{\arraystretch}{1.5}
  \caption{Summary of Maximum Kernel Density Estimates for the precessing analysis of 
  GW231123 at $\fref = 10$ Hz as shown in the bottom panel of \figref{fig:231123_e10_PE}. Bayes factors for the quasi-circular vs eccentric hypothesis, $\bayesecc$, calculated from \Eqnref{eq:bayes} and the nonprecessing vs precessing hypothesis, $\bayesprec$, calculated from \eqnref{eq:bayesprec}. 
  The $e_{10}$ KDE is shown in the bottom panel of \figref{fig:231123_e10_kde}. 
  The $\chi_{1,\perp}$ KDE is shown in \figref{fig:231123_chi1perp_e10_kde}.} 
  \label{tab:231123_e10prec_kde_params} 
\begin{ruledtabular}
\begin{tabular}{c|ccc}
Parameters & $|\chi_{i,z}|< 0.5$ & $|\chi_{i,z}|< 0.9$ & $|\chi_{i,z}|< 0.99$ \\  
  & posterior & posterior & posterior \\
\hline 
$\mc (M_\odot)$ & $160.217_{-47.946}^{+30.468}$ & $166.035_{-48.424}^{+25.115}$ & $154.651_{-37.503}^{+36.563}$  \\
$q$ & $0.934_{-0.330}^{+0.052}$ & $0.948_{-0.317}^{+0.040}$ & $0.939_{-0.320}^{+0.047}$ \\
$e_{10}$ & $0.106_{-0.077}^{+0.144}$ & $0.102_{-0.078}^{+0.121}$ & $0.097_{-0.073}^{+0.124}$\\
$\chieff$ & $0.003_{-0.185}^{+0.230}$ & $0.119_{-0.356}^{+0.274}$ & $0.063_{-0.305}^{+0.383}$ \\
$\chi_{1, \perp}$ & $0.430_{-0.259}^{+0.051}$ & $0.701_{-0.334}^{+0.126}$ & $0.672_{-0.321}^{+0.190}$ \\ 
$\bayesecc$ & -0.19 & -0.27 & -0.25\\
$\bayesprec$ & 11.78 & 11.38 & 10.71\\
\end{tabular}
\end{ruledtabular}
\end{table}

\begin{figure}[h!tbp]
    \includegraphics[width=0.46\textwidth]{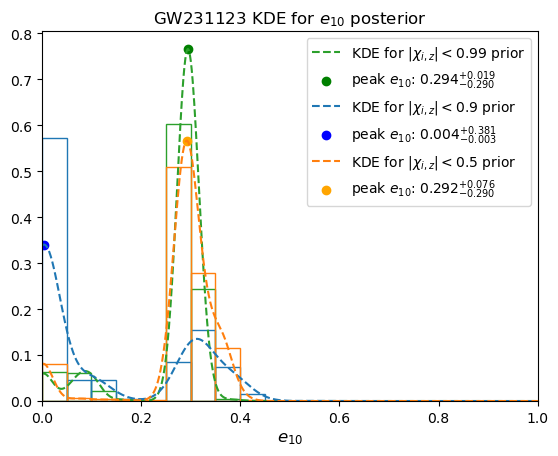}
    \includegraphics[width=0.46\textwidth]{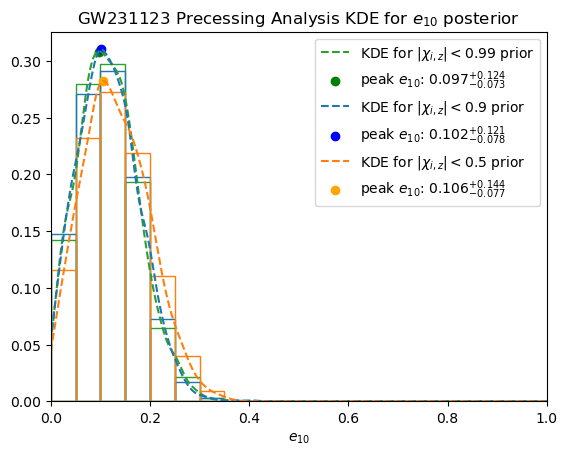}
    \caption{Kernel density estimate of the GW231123 NR-based PE 
    1-dimensional eccentricity posterior at 10 Hz from the nonprecessing analysis 
    ({\it top}) and the precessing analysis ({\it bottom})
    (histograms from \figref{fig:231123_e10_PE}) for the three different spin 
    priors, $|\chi_{i,z}|< 0.99$ (green), $|\chi_{i,z}|< 0.9$ (blue) and 
    $|\chi_{i,z}|< 0.5$ (orange). The maximum of the KDE for each prior is indicated 
    by a dot of the same color with errors reported as the 90\% confidence interval 
    of the histogram.}
    \label{fig:231123_e10_kde}
\end{figure}

\begin{figure}[h!tbp]
    \includegraphics[width=0.46\textwidth]{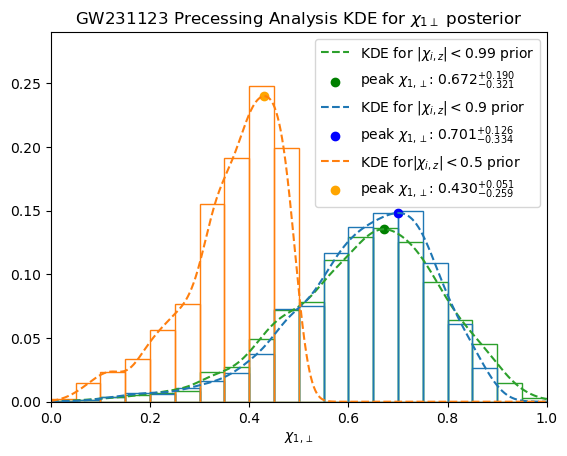}
    \caption{Kernel density estimate of the GW231123 NR-based PE 
    1-dimensional $\chi_{1,\perp}$ posterior at 10\,Hz from the aligned plus 
    precessing spin analysis (histograms from bottom panel of 
    \figref{fig:231123_e10_PE}) for the three different spin priors, 
    $|\chi_{i,z}|< 0.99$ (green), $|\chi_{i,z}|< 0.9$ 
    (blue) and $|\chi_{i,z}|< 0.5$ (orange). The maximum of the KDE for each prior is 
    indicated by a dot of the same color with errors reported as the 90\% confidence 
    interval of the histogram.}
    \label{fig:231123_chi1perp_e10_kde}
\end{figure}


\newpage
\bibliography{combined_bib,references}

\end{document}